\documentclass[11pt]{article}

\usepackage[margin=1in]{geometry}
\usepackage{mathpazo}
\usepackage{subcaption}
\usepackage[dvipsnames]{xcolor}
\usepackage[colorlinks,linkcolor=black,citecolor=OliveGreen,urlcolor=PineGreen]{hyperref}
\usepackage{bbm}
\usepackage[disable]{todonotes}
\usepackage{lineno}
\usepackage[T1]{fontenc}

\usepackage{inconsolata}
\usepackage{url}            
\usepackage{booktabs}       
\usepackage{amsfonts}       
\usepackage{nicefrac}       
\usepackage{microtype}      
\usepackage{multicol}
\usepackage{enumitem}
\usepackage{psfrag}
\usepackage{verbatim}
\usepackage{mathrsfs}
\usepackage{amssymb}%
\usepackage{pifont}%
\usepackage{multirow}
\usepackage{longtable}
\usepackage{listings}
\usepackage{graphicx}
\usepackage{amsmath}
\usepackage{mathtools}
\usepackage{amsthm}
\usepackage{xspace}
\usepackage[subtle]{savetrees}
\usepackage[most]{tcolorbox}
\usepackage{colortbl}
\usepackage[capitalize,noabbrev]{cleveref}

\newcommand\blfootnote[1]{%
  \begingroup
  \renewcommand\thefootnote{}\footnote{#1}%
  \addtocounter{footnote}{-1}%
  \endgroup
}

\makeatletter
\newcommand\BeraMonottfamily{%
  \def\fvm@Scale{0.85}
  \fontfamily{fvm}\selectfont
}
\makeatother
\definecolor{codeerror}{rgb}{0.8,0.1,0.0}
\definecolor{codewarn}{rgb}{0.86,0.3,0.0}

\definecolor{codegreen}{rgb}{0,0.6,0}
\definecolor{codegray}{rgb}{0.5,0.5,0.5}
\definecolor{codepurple}{rgb}{0.58,0,0.82}
\definecolor{backcolour}{rgb}{0.95,0.95,0.95}

\lstdefinestyle{mystyle}{
    backgroundcolor=\color{backcolour},   
    commentstyle=\color{codegreen},
    keywordstyle=\color{magenta},
    numberstyle=\tiny\color{codegray},
    stringstyle=\color{codepurple},
    basicstyle=\BeraMonottfamily\footnotesize,
    breakatwhitespace=false,         
    breaklines=true,                 
    captionpos=b,                    
    keepspaces=true,                 
    numbers=none,                    
    numbersep=5pt,                  
    showspaces=false,                
    showstringspaces=false,
    showtabs=false,                  
    tabsize=2
}

\title{FVEval: Understanding Language Model Capabilities \\ in Formal Verification of Digital Hardware}

\author{%
Minwoo Kang \textsuperscript{1}\blfootnote{Work done as intern at NVIDIA}$^{*}$, 
  Mingjie Liu \textsuperscript{2}, 
Ghaith Bany Hamad \textsuperscript{2}, 
Syed Suhaib \textsuperscript{2}, 
Haoxing Ren \textsuperscript{2}  \\
  \textsuperscript{1} UC Berkeley \textsuperscript{2} NVIDIA\\
  \vspace{0.5mm}\\
  \texttt{mkang@cs.berkeley.edu}\\
  \texttt{\{mingjiel, gbanyhamad, ssuhaib, haoxingr\}@nvidia.com} \\
}
\date{}

\newcommand{\OURS}{FVEval\xspace}

\begin{document}

\maketitle
\begin{abstract}
The remarkable reasoning and code generation capabilities of large language models (LLMs) have spurred significant interest in applying LLMs to enable task automation in digital chip design.
In particular, recent work has investigated early ideas of applying these models to formal verification (FV), an approach to verifying hardware implementations that can provide strong guarantees of confidence but demands significant amounts of human effort.
While the value of LLM-driven automation is evident, our understanding of model performance, however, has been hindered by the lack of holistic evaluation. 
In response, we present FVEval, the first comprehensive benchmark and evaluation framework for characterizing LLM performance in tasks pertaining to FV. 
The benchmark consists of three sub-tasks that measure LLM capabilities at different levels---from the generation of SystemVerilog assertions (SVAs) given natural language descriptions to reasoning about the design RTL and suggesting assertions directly without additional human input.
As test instances, we present both collections of expert-written verification collateral and methodologies to scalably generate synthetic examples aligned with industrial FV workflows. 
A wide range of existing LLMs, both proprietary and open-source, are evaluated against FVEval, based on which we investigate where today’s LLMs stand and how we might further enable their application toward improving productivity in digital FV. 
Our benchmark and evaluation code is available at \url{https://github.com/NVlabs/FVEval}. 
\end{abstract}
\section{Introduction}
\label{sec:introduction}
Hardware verification is a critical stage in digital VLSI development that is not only challenging but also indispensable to successful chip manufacturing.
To ensure that design implementations are fully verified against expected functional behaviors, industrial teams are increasingly adopting \emph{formal} verification (FV), an approach that formulates specifications of the design as temporal logical properties and applies formal methods to mathematically prove that the said properties hold across all expected input stimuli.
While FV offers distinct advantages, such as the ability to detect bugs that are elusive or hard-to-reach with simulation-based approaches, it also bears limitations: FV workflows suffer from steep engineering costs of manually crafting formal testbenches and collateral by human experts.
As a result, scaling the adoption of FV and extending its use to cover a larger number of sub-systems remains a challenge.

With the recent advancements of large language models (LLMs) \cite{Achiam2023GPT4TR, llama3, touvron2023llama, Jiang2024Mixtral, Anil2023Gemini}, researchers have in turn started to apply language models toward automating and improving productivity in hardware verification \cite{Liu2023ChipNeMoDL,He2023ChatEDA,Thorat2023LLMVerilogDev,Tsai2023RTLFixer}.
In particular, recent work has investigated the possibility of LLMs to perform tasks in formal verification such as generating formal assertions from natural language (NL) specifications~\cite{Fang2024AssertLLM} and from register-transfer level (RTL) implementations of the design-under-test (DUT)~\cite{OrenesVera2023autosva2}.
The shared motivation across this recent interest in utilizing LLMs is intuitive: the information that human FV engineers process as input (design RTL and specifications) and the generated outputs (FV testbench implementations) are both innately represented as text and code, with which LLMs have shown impressive capabilities in reasoning and generation \cite{humaneval, hendrycks2021mmlu, Austin2021ProgramSW_MBPP}.

While prior work has demonstrated that current LLMs have the potential to generate hardware assertions, their evaluation, however, has been limited to a hand-select set of test instances with typically less than 10 or 20 problems.
Due to the dearth of publicly available repositories of reliable designs and formal testbenches, it has been challenging to curate a dataset for evaluation that encompasses a variety of designs.
Furthermore, prior evaluation of LLMs has considered a limited set of tasks which renders a holistic evaluation of model capabilities unviable.
This altogether raises the question: how should we \emph{comprehensively} understand and quantify current LLMs capabilities in performing tasks in hardware formal verification?

\begin{figure}[t]
\centering{
\includegraphics[width=0.95\linewidth]{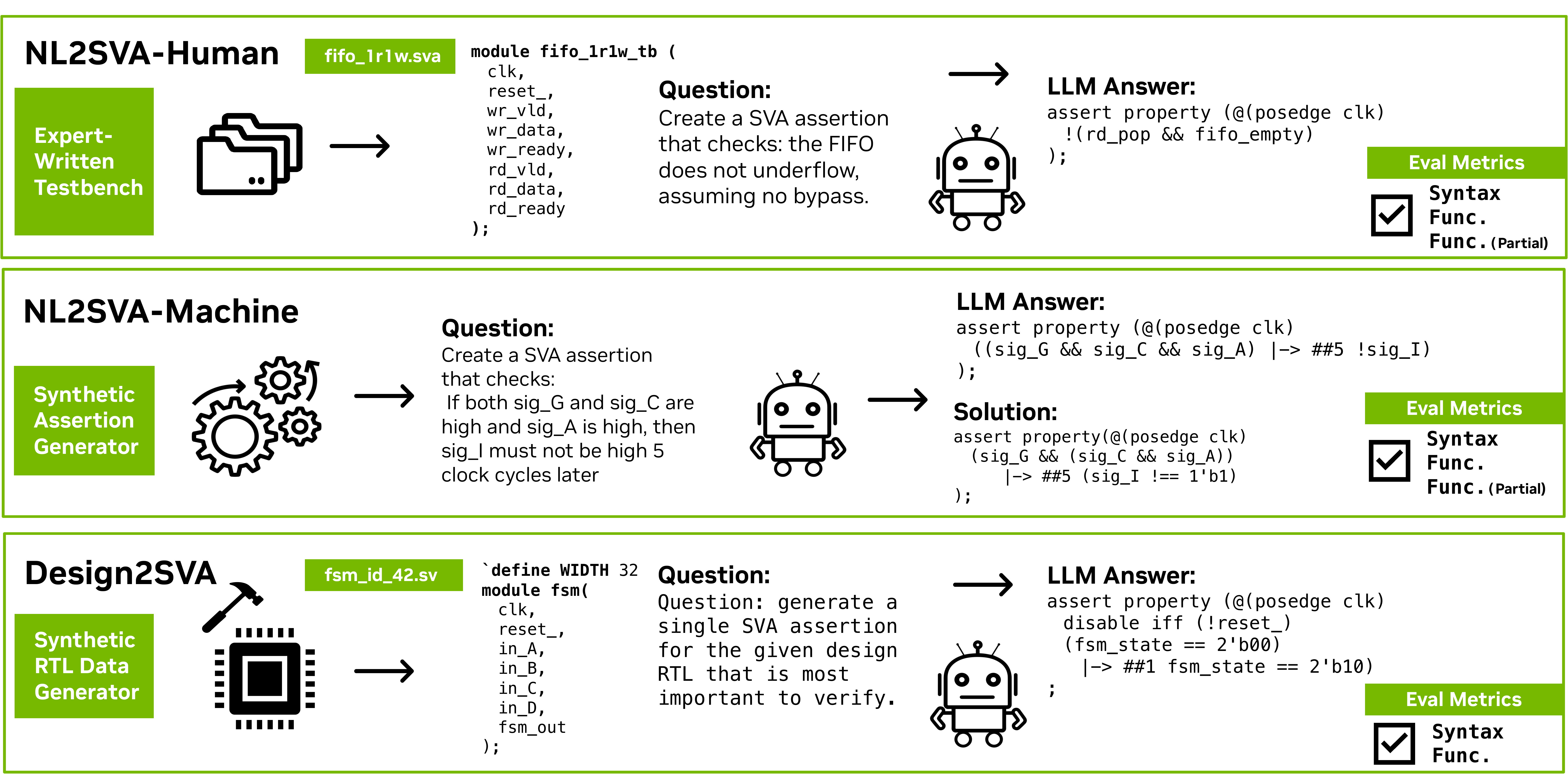}
  \caption{Overview of the \OURS benchmark and evaluation flow. \OURS consists of three sub-benchmarks NL2SVA-Human, NL2SVA-Machine, and Design2SVA that measure model capabilities in generating functionally correct implementations of assertions from NL descriptions and also directly from design under test. The evaluation framework integrates industry-standard FV tools to accurately measure LLM response correctness. }
  \label{fig:overview}
}
\end{figure}

In this work, we present the first comprehensive benchmark for LLMs on diverse tasks pertaining to formal verification via SystemVerilog assertions, which we refer to as \emph{\OURS}. 
As shown in \cref{fig:overview}, \OURS is a collection of three sub-benchmarks: NL2SVA-Human, NL2SVA-Machine, and Design2SVA which each evaluate LLMs in different industry-relevant scenarios exercising various capabilities related to code generation and reasoning in the context of SystemVerilog and hardware formal properties.
As test problem instances, we either collect human-written designs and testbenches or present methodologies to scalably generate synthetic yet realistic test cases motivated by real-world designs. 
Finally, we evaluate a wide range of existing LLMs, both proprietary and open-source, against \OURS and delineate the limitations of current models.
\section{Preliminaries}
\label{section:preliminaries}

Hardware verification is a crucial stage in the chip design process, necessary to make sure the end product functions according to the specification. However, even though on average more resources are now allocated to verification than hardware design, recent industry data shows that an increasing percentage of projects see critical bugs escape verification \cite{foster2022} that must be fixed by an expensive re-spin. 
There are two widely accepted approaches to hardware verification: traditional simulation-based verification and formal verification (FV), which has been gaining momentum in the recent decade. 

Unlike simulation-based verification, FV aims to formally prove that a digital circuit design meets a given specification \cite{kropf1999introduction}. The specification in FV is represented as a set of properties that either constrain the input stimuli, as \textit{assumptions}; or define the expected behavior of the design to be verified, as \textit{assertions}. Digital circuits can naturally be expressed as state transition systems, and the properties in temporal logic \cite{ClarkeEmerson86, Emerson82}. This representation allows model checkers to traverse through the state transition graph to search for any states where the properties do not hold \cite{clarke2001bounded}. If the model checker can find any such state, it returns a \textit{counterexample} or trace for which that property fails. In contrast, simulation-based verification would only check the properties for a subset of states based on the defined input stimuli. If a model checker can provide a formal guarantee that a property is indeed true in \textit{every} state---it has found a formal \textit{proof} the property always holds for the design.

\section{Overview of the \OURS Benchmark Framework}
\label{section:overview}
This section presents an overview of the proposed \OURS---a collection of three sub-benchmarks that assess language models on varying tasks related to hardware FV.
Here, we first describe the motivation for formulating these sub-benchmarks and the capabilities of LLMs that we seek to measure.
Then we describe the details, including the dataset collection or generation process, definitions of model inputs and outputs, and evaluation metrics for each sub-benchmark.

\subsection{Motivation}
In this work, we focus on assessing the viability of LLMs to be applied towards industrial hardware FV based on SystemVerilog Assertions (SVAs).  
In performing functional verification, FV engineers go through the process of: (1) analyzing the design-under-test (DUT) RTL and the specification documentation describing the expected functional behaviors of the design; and (2) writing testbenches that consist of assertions checking the design against the specification. 
LLMs could potentially improve FV productivity by automating the implementation of such SVA assertions from natural language specifications. 

\begin{figure}[b!]
\centering{
\includegraphics[width=0.85\linewidth]{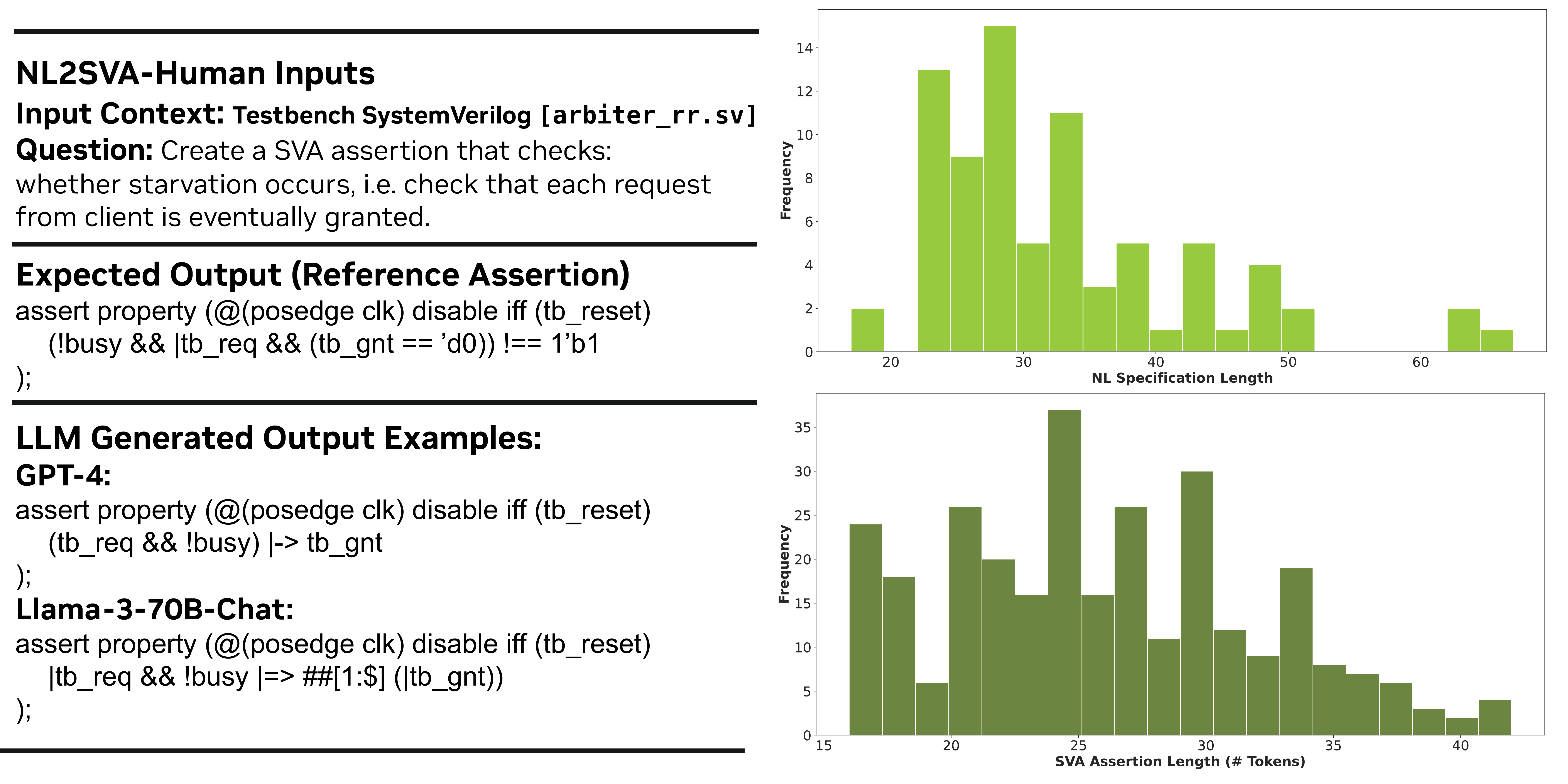}
\caption{(Left) Input and output examples of the NL2SVA-Human benchmark. 
Reference solution is the SVA assertion written by the human engineer that the LLM response is expected to match in terms of functionality.
(Right) Distribution of token length for the NL specifications and the reference SVA solutions contained in the benchmark. }
\label{fig:nl2sva_human_examples}
}
\end{figure}

In this context, we curate the first sub-benchmark, \textbf{NL2SVA-Human}, through which we raise the question: \emph{Do LLMs have the capability to generate SVA assertions, given real-world, human-written testbenches and high-level specifications of design functionality?}
Even though SVA assertions can be relatively short pieces of code, the outputs must accurately reflect the specification and do so in the context of the provided testbench, which entails grounding the implementation of the assertion on existing modeling code provided in the testbench.
In curating the evaluation dataset, we focus on a representative set of testbenches that target common unit-level modules that  engineers repeatedly encounter as part of larger sub-systems. 
Therefore, models that can reliably generate SVA assertions for test instances in NL2SVA-Human may have great potential for industrial utility.

We also note that in practice, FV engineers not only repeat similar patterns of assertions for common sub-units but also implement custom assertions tailored to a specific property to be checked. 
\textbf{NL2SVA-Machine} attempts to quantify LLM performance the second case: \emph{"Can LLMs flexibly handle diverse NL specifications of formal properties and accurately formulate the same formal logic in SVA syntax?"}
The emphasis here is on evaluating LLMs against a wide range of NL to SVA assertion pairs that exercise various formal logical operators.
High accuracy on this benchmark reflects the robustness of a model in generating code outputs in SystemVerilog Assertions.

Finally, a more challenging but also extremely helpful application of LLMs is their use towards suggesting formal properties to be verified for a particular design RTL and prior to a human engineer drafting a full specification.
In \textbf{Design2SVA}, we ask: \emph{"Can LLMs craft relevant SVA assertions directly from design RTL and without human guidance?"}
Compared to previous benchmarks, this task demands a higher-level of understanding RTL code and the semantics of the hardware module described by the RTL.
Models successful in this benchmark could have potential to realize further advanced usages of language models in hardware FV, as artificial agents assisting human engineers.

\subsection{NL2SVA-Human: Assertion Generation with Real-World Testbenches}
\textbf{Dataset.} 
Our collection of test instances cover basic units of FV such as first-in-first-out (FIFO) queues, arbiters, hardware counters, random-access memory (RAM) units, and finite-state machines.
Each testbench comes with a set of assertions and accompanying natural language description written by formal experts that address various design functionalities, with the benchmark holding a total of 79 assertions.
\cref{fig:nl2sva_human_examples} summarizes the statistics about the test instances, in terms of the distribution of lengths of natural language specifications and reference SVA assertion solutions.
We measure length as the number of tokens based on the tokenizer used in the Llama3 models \cite{llama3}.
We see a wide range in the distribution reflecting the variety of formal properties in the benchmark.

Models are given as prompt input: (1) the testbench code (SystemVerilog) with all of its original assertions removed but containing modeling code with internal signals (wires and state) and input/output ports of the testbench module; and (2) a \emph{high-level} NL description of the intended assertion.
An example is shown on the left of \cref{fig:nl2sva_human_examples}.
The model response is expected to match the human-written assertion that serves as ground-truth.

\paragraph{Evaluation.} 
We evaluate model responses in terms of syntax and functional correctness.
For syntax correctness, we measure whether the generated SVA assertions pass the syntax check performed by industry-standard, commercial FV tools such as Cadence Jasper.
For functional correctness, we present a novel approach in measuring \emph{exact logical equivalence} between a pair of assertions, model-generated and ground-truth.
Our method utilizes a custom-implemented Jasper function that \emph{formally} proves logical equivalence, and further, can also evaluate whether there is an implication relationship between the two assertions.
We also report a relaxed metric of functional accuracy that includes such a case of \emph{partial equivalence} and thereby capture the performance of LLMs at a finer granularity than only considering exact equivalence.
Finally, as an alternative measure of functional correctness, the lexical similarity between the model-generated code against the reference solution using standard $n$-gram metrics such as BLEU could be considered.
While we report BLEU scores, we do note that commercial FV tools are required for accurate evaluation of SVA assertions.

\begin{figure}[t!]
\centering{
\includegraphics[width=0.85\linewidth]{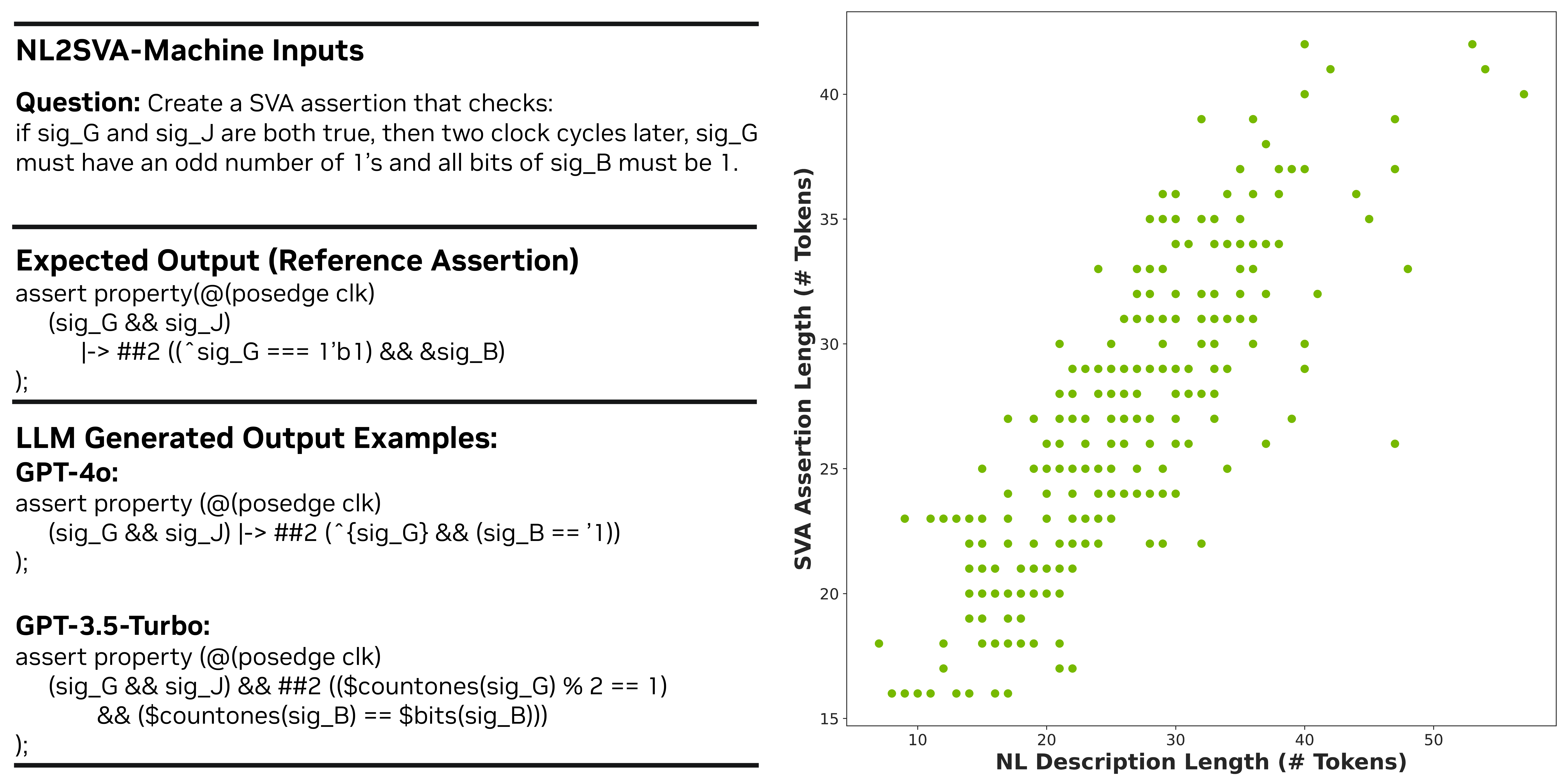}
\caption{(Left) Input and output examples of the NL2SVA-Machine benchmark. 
The dataset consists of naturalized descriptions of the formal-symbolic logic formulae and the corresponding SVA assertion, based on which the NL annotations were created.
(Right) Distribution of lengths of NL descriptions and SVA assertions, measured in token length using the Llama3 tokenizer.}
\label{fig:nl2sva_machine_examples}
}
\end{figure}

\subsection{NL2SVA-Machine: Synthetic Benchmark for Stress-Testing Formal Assertion Generation}
\paragraph{Dataset.} 
To create diverse test cases of NL description and SVA assertion pairs, we follow the process of: 
(1) random SVA assertion generation, based on random sampling of SVA operators and symbolic signal names;
(2) LLM generation of NL descriptions for each random assertion;
(3) LLM as a critic to assess whether the NL description accurately reflects the temporal logic of the assertion---if this fails, re-try description generation;
and (4) Human inspection to finalize the appropriateness of generated descriptions.
For description generation, we run LLM inference with sampling temperature $T=1.0$ and for assessment as a critic, we use greedy sampling with temperature $T=0.0$.
We use \texttt{gpt-4o} for NL description generation and \texttt{gpt-4-turbo} for assessment.

With a fully automated generated flow, NL2SVA-Machine consists of 300 test cases with widely varying patterns of SVA assertions and natural language usage variations to describe them.
An example of a test instance and sample LLM responses are shown in \cref{fig:nl2sva_machine_examples}.
As input context, models are solely given a symbolic text description of the logical formula, and the expected output is an SVA assertion matching the logical expression. 
Summaries of the distribution of randomly generated test cases are shown in \cref{fig:nl2sva_machine_examples}.

\paragraph{Evaluation.}
As NL2SVA-Machine also prompts models with natural language specifications and expects outputs in concrete SVA assertion implementations, we follow the same evaluation protocol and metrics as NL2SVA-Human.

\subsection{Design2SVA: Direct Generation of Formal Assertions from Design RTL Alone}

\paragraph{Dataset.}
Our objective in formulating the Design2SVA benchmark is to present a set of test cases consisting of design RTL that are: 
(1) sufficiently correct, as in there exist formal properties that can be proven to be true; (2) relevant to real-world FV use-cases; (3) suitable as test instances for language models; and (4) varied in terms of complexity.
While it is ideal to collect and curate existing RTL examples, we find that openly available SystemVerilog/Verilog repositories contain limited numbers of verified RTL designs.
Of the cases that satisfy the first and second criteria, such as module instances from OpenTitan~\cite{opentitan}, these cases fail to be suitable for language model evaluation, as each module is a part of a large System-on-a-Chip (SoC) system and the context needed to resolve all sub-module dependency information is prohibitively large.
\begin{figure}[t!]
\centering{
\includegraphics[width=0.85\linewidth]{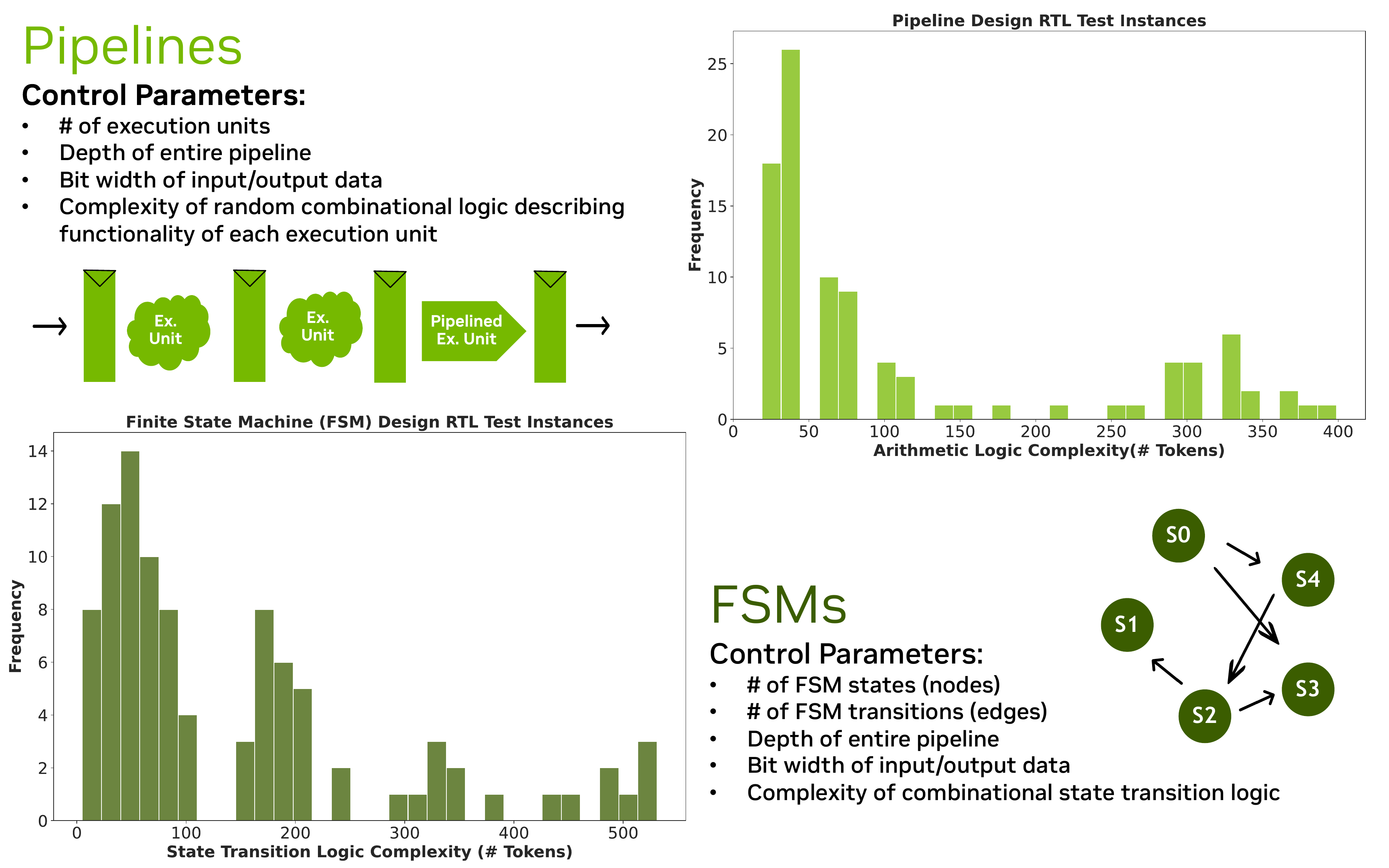}
  \caption{Synthetic RTL Testcase generation. For each category of designs, we vary a list of control parameters to generate test design instances such that the total set of 96 cases for each category exhibit a wide distribution of difficulties. The total token length of the randomly generated arithmetic logic and FSM transition logic, correlated with complexity of designs under FV, are shown.}
  \label{fig:synthetic_designs}}
\end{figure}
\begin{figure}[b!]
\centering{
\includegraphics[width=0.75\linewidth]{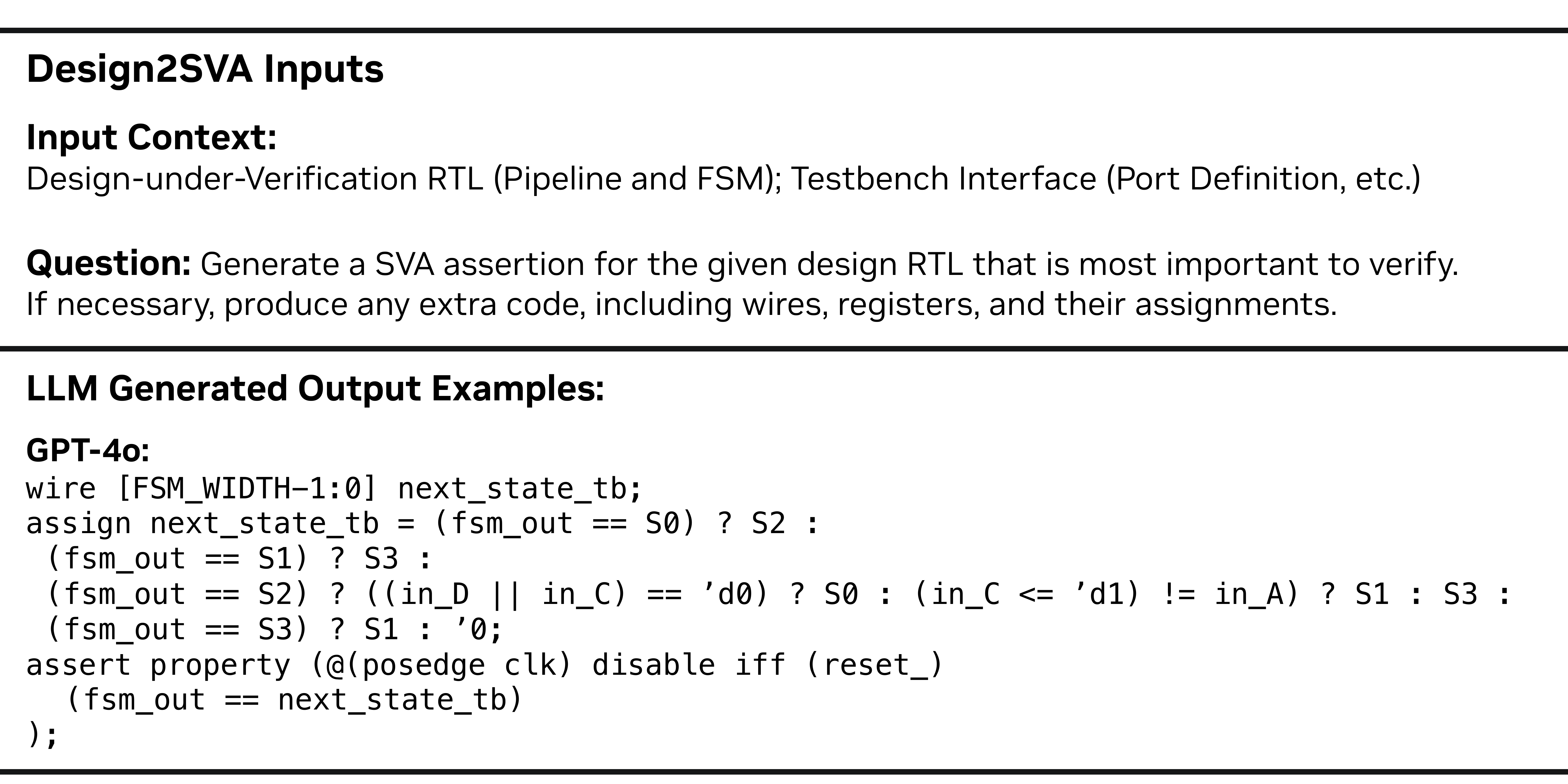}
\caption{Input and output examples of the Design2SVA benchmark.
Models are provided with the design RTL as input context, based on which are then prompted to generate valid assertions.
The model is also prompted to generate any accompanying code necessary for implementing the given assertion, as shown in the example response.}
}
\label{fig:design2sva-examples}
\end{figure}
Instead, we propose a methodology to scalably generate complex and parameterized synthetic test instances that are derived from common design patterns encountered in industrial FV workflows.
As shown in \cref{fig:synthetic_designs}, the two categories of RTL designs we generate are (1) arithmetic pipelines that resemble scenarios where we check for data integrity and forward propagation across datapaths; and (2) finite-state machines (FSMs) that commonly appear in control logic implementations, such as cache controllers, memory interfaces, etc.
Each category of our designs has randomized sub-components---the arithmetic execution units in the pipeline designs and the FSM graph and state transition logic in the FSM design---and each generated RTL is parameterized such that controlled generation of test instances is possible.
While a virtually limitless number of designs can be generated in this manner, we compose the Design2SVA benchmark to contain 96 test instances for each design category (pipeline and FSM) based on a controlled sweep of generator parameters.

\paragraph{Evaluation}
For each generated RTL test instance, we also generate an accompanying formal testbench, based on which language models generate suggestions of assertions. 
The \OURS evaluation flow then re-formats the testbench with the model-generated assertions and supplies the SystemVerilog to the commercial tool backend for evaluation.
The metrics of evaluation are similar to the NL2SVA benchmarks:
(1) Syntax---we measure the LLM generated assertion is first syntactically correct;
(2) Functionality---we use the results of formal proofs, \emph{i.e.} whether the assertions are proven with model checkers and other formal engines in industrial tools, as an indication of functional correctness.

Unlike in NL2SVA-Human and NLSVA-Machine, the Design2SVA benchmark subjects LLMs to generate relevant and correct SVA assertions given a design under verification, and as such, there are numerous valid completions that are functionally different from each other.
To quantify model capability in this setting, we consider the \emph{pass@k} metric, where a problem instance is considered solved if any of the $k$ attempted solutions are correct.
\begin{figure}[ht]
    \centering
    \begin{minipage}{0.5\textwidth}
        \centering
        \resizebox{\textwidth}{!}{%
            \begin{tabular}{l|cccc}
                \toprule
                \textbf{Model} & \texttt{Syntax} & \texttt{Func.} & \texttt{Partial Func.}& BLEU \\
                \midrule
                gpt-4o & \underline{0.911} & \textbf{0.456} & \textbf{0.582} & 0.503 \\
                gemini-1.5-pro & 0.810 & 0.253 & 0.380 & 0.484 \\
                gemini-1.5-flash & \textbf{0.949} & \underline{0.380} & \underline{0.557} & 0.518 \\
                \midrule
                Mixtral-8x22b & 0.823 & 0.190 & 0.278 & 0.450  \\
                Llama-3.1-70b & 0.861 & 0.291 & 0.354 & 0.464 \\
                Llama-3-70b & 0.899 & 0.291 & 0.506 & 0.464  \\
                Llama-3.1-8b & 0.835 & 0.203 & 0.304 & 0.525 \\
                Llama-3-8b & 0.747 & 0.063 & 0.215 & 0.491 \\
                \bottomrule
            \end{tabular}
        }
        \captionof{table}{Evaluation results for the NL2SVA-Human Benchmark. Boldface and underlined results indicate highest and second highest metric values, respectively. We see that models that proprietary models with generally advanced capabilities for code generation and logical reasoning achieve the best performance across all metrics. Notably, all models show a significant gap between full and \emph{partial} equivalence of assertion functionality---this highlights the value of fine-grained measurement of functional correctness that captures how close LLM-generated assertions are similar to the human-written reference solution.}
        \label{tab:nl2sva_human_result}
    \end{minipage}%
    \hfill
    \begin{minipage}{0.45\textwidth}
        \centering
        \includegraphics[width=\textwidth]{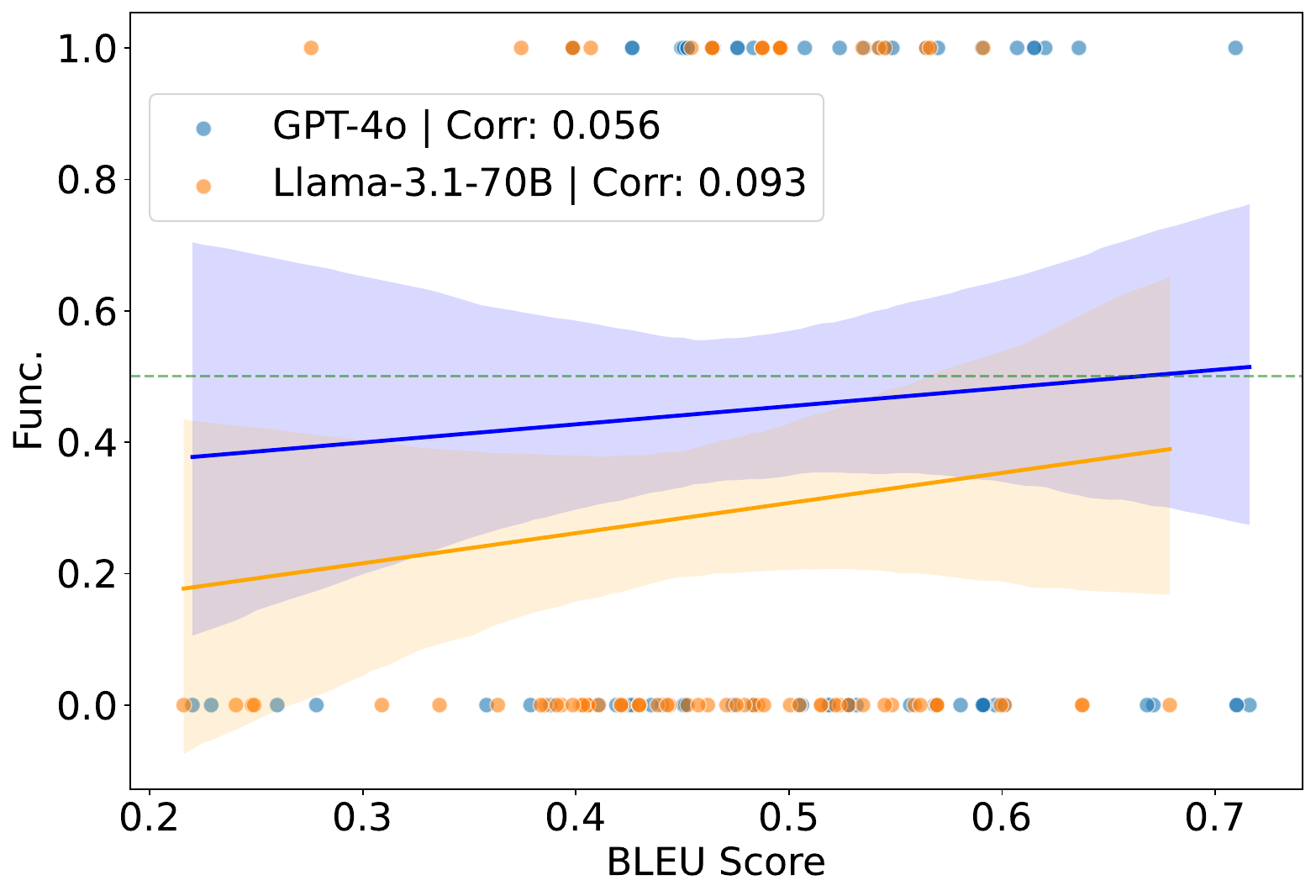}
        \vspace{-1.5em}
        \caption{Analysis on the correlation between functionality correctness measures based on formal equivalence versus BLEU scores as approximate measures of assertion similarity. We see little correlation between the two metrics, with coefficients 0.0563 and 0.0926 respectively, indicating that BLEU scores are insufficient to fully capture the equivalence of a LLM-generated assertion against a reference solution.}
        \label{fig:nl2sva_human_bleu_score_analysis}
    \end{minipage}
\end{figure}

\section{Results and Analysis}
\label{section:results}
In this section, we report the results of evaluating a suite of current state-of-the-art language models, trained on considerable sizes of text corpora pertaining to diverse programming languages.

\subsection{Experiment Setup}
A wide range of LLMs, both proprietary and open-source (i.e. models with weights publicly available) are considered.
We consider the most recent OpenAI model \texttt{gpt-4o-0513} \cite{Achiam2023GPT4TR} and Google Gemini models \cite{Anil2023Gemini} \texttt{gemini-1.5-pro-001} and \texttt{gemini-1.5-flash-001}.
All results of the proprietary models are based on accessing their respective endpoint services in May 2024.
For open-source models, we evaluate the models from the Llama 3 \cite{llama3} and Llama 3.1 family \cite{llama3p1} with 8B and 70B parameters; we also consider \texttt{mixtral-8x22b-instruct-v1}, a sparse Mixture-of-Experts model with approximately 39B active parameters \cite{Jiang2024Mixtral}.
Note that all models are instruction fine-tuned models, as our tasks heavily require instruction-following capabilities.
Evaluation of open-source models was performed via vLLM \cite{kwon2023efficient}, a high-throughput LLM inference framework. 

\subsection{NL2SVA-Human Results}

We consider zero-shot model inference where (1) the problem testbench and (2) NL descriptions of the intended SystemVerilog assertion is given as model input.
Evaluation results from running model inference with greedy decoding are summarized in \cref{tab:nl2sva_human_result}. 
We observe that the models reported with highest general capabilities in code generation and logical reasoning, such as \texttt{gpt-4o} and \texttt{gemini} models, show the best performance across all metrics.

\begin{tcolorbox}[width=1.0\linewidth, halign=left, colframe=black, colback=white, boxsep=0.01mm, arc=1.5mm, left=2mm, right=2mm, boxrule=1pt]
\footnotesize{
\textbf{Problem ID: \texttt{fifo\_1r1w\_bypass\_4}}\\
\vspace{1em}
Question: Create a SVA assertion that checks: that when there is a write push to the FIFO, data is eventually popped. Use the signals 'rd\_pop' and 'wr\_push'.

\vspace{1em}

\textbf{Reference Solution:}
\begin{lstlisting}[language=Verilog,  basicstyle=\BeraMonottfamily\scriptsize]
asrt: assert property (@(posedge clk) disable iff (tb_reset)
    wr_push |-> strong(##[0:$] rd_pop)
);
\end{lstlisting}

\textbf{\texttt{gpt-4o} Response:}\\
\begin{lstlisting}[language=Verilog,  basicstyle=\BeraMonottfamily\scriptsize]
asrt: assert property (@(posedge clk) disable iff (tb_reset)
    wr_push |-> ##[1:$] rd_pop // weak temporal implication + one-cycle latency
); // reference assertion implies this assertion
\end{lstlisting}
Syntax: \textcolor{blue}{pass} | Functionality: \textcolor{YellowOrange}{partial pass} \\

\vspace{1em}

\textbf{\texttt{Llama-3.1-70B-Instruct} Response:}\\
\begin{lstlisting}[language=Verilog,  basicstyle=\BeraMonottfamily\scriptsize]
asrt_wr_push_rd_pop: assert property (@(posedge clk) disable iff (tb_reset)
    wr_push |-> eventually(rd_pop)
); // 'eventually' is an invalid operator
\end{lstlisting}
Syntax: \textcolor{red}{fail} | Functionality: \textcolor{red}{fail} \\
\vspace{1em}

\hrule
\vspace{1em}

\textbf{Problem ID: \texttt{arbiter\_reverse\_priority\_9}}
\vspace{1em}

Question: Create a SVA assertion that checks: that the arbiter is never on hold or busy or on continued grant at the same time. Use the signals 'busy', 'hold', and 'cont\_gnt'.
\vspace{1em}

\textbf{Reference Solution:}
\begin{lstlisting}[language=Verilog,  basicstyle=\BeraMonottfamily\scriptsize]
asrt: assert property (@(posedge clk) disable iff (tb_reset)
    !$onehot0({hold,busy,cont_gnt}) !== 1'b1
);
\end{lstlisting}

\textbf{\texttt{gpt-4o} Response: }\\
\begin{lstlisting}[language=Verilog,  basicstyle=\BeraMonottfamily\scriptsize]
asrt: assert property (@(posedge clk) disable iff (tb_reset)
    !(busy && hold && cont_gnt) // this assertion only checks if all three signals are set high
); // this assertion implies the reference
\end{lstlisting}
Syntax: \textcolor{blue}{pass} | Functionality: \textcolor{YellowOrange}{partial pass} \\

\vspace{1em}
\textbf{\texttt{Llama-3.1-70B-Instruct} Response:}\\
\begin{lstlisting}[language=Verilog,  basicstyle=\BeraMonottfamily\scriptsize]
asrt: assert property (@(posedge clk) disable iff (tb_reset)
    !(busy && (hold || cont_gnt)) && !(hold && (busy || cont_gnt)) && !(cont_gnt && (busy || hold))
);
\end{lstlisting}
Syntax: \textcolor{blue}{pass} | Functionality: \textcolor{blue}{pass} \\
}
\end{tcolorbox}
\noindent\begin{minipage}{\textwidth}
\captionof{figure}[hypcap=false]{Examples of common failure modes observed in different language models for the NL2SVA-Human benchmark. (1) LLMs frequently hallucinate SVA syntax and operators, such as ``s\_eventually''. (2) Given a high-level natural language specification, models may generate assertions that either imply or is implied by the reference solution: while these assertions are not exactly equivalent to the intended assertions, it is of interest to capture such cases of proximal solutions when evaluating different LLMs' applicability to the use case of generating SV assertions from high-level natural language descriptions.}\label{box:nl2sva_human_failure_modes}
\end{minipage}
\begin{table}[ht]
\centering
    \resizebox{0.7\textwidth}{!}{%
    \begin{tabular}{l|ccccc}
    \toprule
    \textbf{Model} & \texttt{Syntax@5} & \texttt{Func.@3} & \texttt{Func.@5} & \texttt{Partial.@3} & \texttt{Partial.@5} \\
     \midrule
gpt-4o & 0.987 & 0.461 & 0.468 & 0.614 & 0.651 \\
gemini-1.5-flash & 0.987 & 0.442 & 0.466 & 0.602 & 0.629 \\
Llama-3.1-70b & 0.975 & 0.382 & 0.436 & 0.545 & 0.604 \\
    \bottomrule
    \end{tabular}
}
\caption{Results for NL2SVA-Human under the case of sampling multiple model responses, where \texttt{pass@k} for each evaluation metric is measured. We observe that all models are able to generate at least one syntactically correct SVA implementation out of five attempts; functional correctness is also improved with increasing numbers of samples.}
\vspace{-1em}
\label{tab:nl2sva_human_result_with_sampling}
\end{table}

\textbf{LLMs can generate syntactically correct SVA code but are still prone to hallucinations.} Even the most capable models considered, we observe that models are still prone to hallucinating subsets of SVA syntax, e.g. \texttt{s\_eventually}, \texttt{strong}, and other particular operators. In \cref{box:nl2sva_human_failure_modes}, we see that \texttt{Llama-3.1-70B} incorrectly applying a non-existent operator ``eventually'' in generating a liveness assertion, likely due to the key word provided in the natural language input. Such hallucination is also observed in the most capable models including \texttt{gpt-4o}, \texttt{gemini-1.5-pro}, and \texttt{gemini-1.5-flash}. 

\textbf{Partial functional equivalence as a correctness metric reveals further understanding of model capabilities.} 
Overall, there is a notable gap between the accuracy measured with exact functional equivalence and that with partial equivalence.
Such a gap can be attributed to the innate ambiguity in the high-level NL descriptions---while full functional equivalence necessitates correctness, those that imply the ground-truth solution and vice-versa are also possibly valid solutions given the input descriptions. 
NL2SVA-Human reflects scenarios in the industrial FV workflows, where engineers wish to benefit from LLM-based auto-generation of SVA assertions from possibly ambiguous NL specifications.
In the examples from \cref{box:nl2sva_human_failure_modes}, we indeed see that partially equivalent assertions are reasonable responses to the natural language input, and taking into account both the full and partial functional equivalence of LLM-generated solutions gives a richer picture of how these models would be applicable in scenarios where engineers provide high-level sketches of desired assertion functionality.
\OURS is the first evaluation framework that elucidates analysis on this nature of completing formal assertion implementation from natural language, and is able to provide concrete metrics of full and partial equivalence based on industry-standard verification tools.

\begin{table}[ht]
    \centering
\resizebox{0.85\textwidth}{!}{%
    \begin{tabular}{l|cccc|cccc}
    \toprule
    \multirow{2}{*}{\textbf{Model}} & 
    \multicolumn{4}{c|}{\textbf{0-shot}} & \multicolumn{4}{c}{\textbf{3-shot}}\\
      & \texttt{Syntax} & \texttt{Func.} & \texttt{Partial Func.}& BLEU 
& \texttt{Syntax} & \texttt{Func.} & \texttt{Partial Func.}& BLEU\\
     \midrule
gpt-4o & \textbf{0.927} & \textbf{0.430} & \textbf{0.540} & 0.622 & \textbf{0.937} & \textbf{0.467} &\textbf{0.570} & 0.660 \\
gemini-1.5-pro & 0.467 & 0.137 & 0.203 & 0.510  & 0.880 & 0.417 & 0.517 & 0.622 \\
gemini-1.5-flash & 0.783 & \underline{0.377} & 0.470 & 0.574 & 0.837 & 0.397 & 0.480 & 0.613 \\
\midrule
Mixtral-8x22b & \underline{0.913} & 0.327 & 0.500 & 0.546 & 0.880 & 0.430 & 0.523 & 0.599 \\
Llama-3.1-70b & 0.887 & 0.303 & 0.397 & 0.595 & \underline{0.920} & \underline{0.457} & \underline{0.567} & 0.651 \\
Llama-3-70b & 0.863 & 0.330 & 0.430 & 0.570 & 0.860 & 0.380 & 0.503 & 0.621 \\
Llama-3.1-8b & 0.813 & 0.320 & \underline{0.520} & 0.533 & 0.840 & 0.267 & 0.370 & 0.582 \\
Llama-3-8b & 0.673 & 0.187 & 0.320 & 0.493 & 0.827 & 0.240 & 0.397 & 0.552 \\
    \bottomrule
    \end{tabular}%
}
\caption{Evaluation results for the NL2SVA-Machine Benchmark. Boldface and underlined results indicate highest and second highest metric values, respectively. While even the smaller models achieve moderately high syntax correctness (>80\%) with in-context examples, we see a limitations in functional correctness and again a gap between full and partial correctness of model-generated assertions.}
\label{tab:nl2sva_machine_result}
\end{table}

\textbf{BLEU scores are not indicative of formal, functional correctness for NL2SVA-Human.} In \cref{fig:nl2sva_human_bleu_score_analysis}, we see that BLEU scores, measuring similarity between LLM outputs against ground-truth assertion implementations, are not correlated with formal equivalence based measures of correctness. While BLEU scores or other lexical measures capture surface-level similarities between outputs, we find that for a complete analysis of LLM capabilities, it is necessary to analyze formal equivalence between assertions, which is implemented as part of \OURS evaluation framework. 

\textbf{Sampling multiple response candidates improves accuracy.} Here, we investigate a scenario where models generate multiple responses in parallel for each test instance. \cref{tab:nl2sva_human_result_with_sampling} presents the results from the best performing LLMs from the zero-shot greedy decoding where we measure the \texttt{pass@5} values of each evaluation metric. We observe that state-of-the-art models, such as \texttt{gpt-4o}, achieve a relaxed functional \texttt{pass@5}, showing an improvement of up to 11\% compared to \texttt{pass@1}. While the absolute success rate is modest, these improvements suggest the potential for future research into novel training and inference strategies, including prompt optimization and inference-time compute scaling \cite{snell2024scaling}, to enhance model generation of SystemVerilog assertions.

\subsection{NL2SVA-Machine Results}

A further task to quantify model capability in handling diverse SystemVerilog operator syntax and temporal logic reasoning based on such grammar, NL2SVA-Machine evaluation examines whether LLMs can accurately reason about the temporal logic expressed in natural language and translate into SVA syntax.
\cref{tab:nl2sva_machine_result} summarizes model performances in zero-shot and with 3-shot inference settings, where the hand-crafted in-context examples are fixed across test questions.
\begin{tcolorbox}[width=1.0\linewidth, halign=left, colframe=black, colback=white, boxsep=0.01mm, arc=1.5mm, left=2mm, right=2mm, boxrule=1pt]
\footnotesize{
\textbf{Problem ID: \texttt{nl2sva\_machine\_3\_61\_0}}\\
\vspace{1em}
Question: Create a SVA assertion that checks: When either sig\_D is high or sig\_H has an odd number of bits set to '1', and sig\_F is also high, the assertion is satisfied.

\vspace{1em}

\textbf{Reference Solution:}
\begin{lstlisting}[language=Verilog,  basicstyle=\BeraMonottfamily\scriptsize]
assert property(@(posedge clk)
    ((sig_D || ^sig_H) && sig_F)
);
\end{lstlisting}

\textbf{\texttt{gpt-4o} | 0-Shot Inference:}\\
\begin{lstlisting}[language=Verilog,  basicstyle=\BeraMonottfamily\scriptsize]
assert property (@(posedge clk)
    (sig_D || ($countones(sig_H) % 2 == 1)) |-> sig_F 
); // sig_F may not be high at the same cycle as sig_D or (sig_H having odd number of bits high)
// for the assertion to hold; thus, this assertion implies the reference solution
\end{lstlisting}
Syntax: \textcolor{blue}{pass} | Functionality: \textcolor{YellowOrange}{partial pass} \\

\vspace{1em}

\textbf{\texttt{gpt-4o} | 3-Shot Inference:}\\
\begin{lstlisting}[language=Verilog,  basicstyle=\BeraMonottfamily\scriptsize]
assert property(@(posedge clk)
    ((sig_D || (^sig_H)) && sig_F)
);
\end{lstlisting}
Syntax: \textcolor{blue}{pass} | Functionality: \textcolor{blue}{pass} \\

\vspace{1em}
\textbf{\texttt{Llama-3.1-8B-Instruct} | 0-Shot Inference:}\\
\begin{lstlisting}[language=Verilog,  basicstyle=\BeraMonottfamily\scriptsize]
assert property (@(posedge clk)
    (sig_D || ($countones(sig_H) % 2 == 1)) && sig_F
);
\end{lstlisting}
Syntax: \textcolor{blue}{pass} | Functionality: \textcolor{blue}{pass} \\

\vspace{1em}
\textbf{\texttt{Llama-3.1-8B-Instruct} | 3-Shot Inference:}\\
\begin{lstlisting}[language=Verilog,  basicstyle=\BeraMonottfamily\scriptsize]
assert property(@(posedge clk)
    ((sig_D || ($bits(sig_H) % 2 == 1)) && sig_F)
); // '$bits' checks the width of the signal; this assertion implies the reference solution
\end{lstlisting}
Syntax: \textcolor{blue}{pass} | Functionality: \textcolor{YellowOrange}{partial pass} \\

}
\end{tcolorbox}
\noindent\begin{minipage}{\textwidth}
\captionof{figure}[hypcap=false]{Examples of failure modes observed with the NL2SVA-Machine benchmark. While syntactically correct, the assertion implementation from \texttt{gpt-4o} under zero-shot inference is not fully equivalent to the reference solution in relations to the signal \texttt{sig\_F}. However, in the case of \texttt{Llama-3.1-8B} we see the fixed in-context examples distracting the model and leading to incorrectly applying the \texttt{\$bits} operator.}\label{box:nl2sva_machine_failure_modes}
\end{minipage}

\textbf{LLMs achieve high accuracy in terms of syntax but struggle with generating SVA matching the assertion specification expressed as natural language input.}
With in-context learning, models with high parameter count reach close to perfect accuracy in terms of syntax; however, the accuracy on functionality both exact (full equivalence) and relaxed (partial equivalence) does not improve significantly.
The gap between syntax and functional correctness in state-of-the-art models indicate the limitations of these models in accurately identifying the propositional and modal temporal logic expressed in natural language and translating into SVA implementation.

\begin{table}[ht]   
\centering
    \resizebox{0.7\textwidth}{!}{%
    \begin{tabular}{l|ccccc}
    \toprule
    \textbf{Model} & \texttt{Syntax@5} & \texttt{Func.@3} & \texttt{Func.@5} &  \texttt{Partial.@3} & \texttt{Partial.@5} \\
     \midrule
gpt-4o & 0.960 & 0.495 & 0.512 & 0.615 & 0.636 \\
gemini-1.5-flash & 0.940 & 0.462 & 0.483 & 0.556 & 0.583 \\
Llama-3.1-70b  & 0.960 & 0.539 & 0.566 & 0.642 & 0.669 \\
    \bottomrule
    \end{tabular}
}
\caption{Results for NL2SVA-Machine with sampling multiple model responses, where \texttt{pass@k} for each evaluation metric is measured. Models are prompted with same three in-context examples, and we use nucleus sampling with top $p=0.95$ and temperature $t=0.8$. Again, we observe that all models are able to generate at least one syntactically correct SVA implementation out of five attempts; functional correctness is also improved with increasing numbers of samples.}
\label{tab:nl2sva_machine_result_with_sampling}
\end{table}

\textbf{In-context examples can aid models to generate functionally correct implementations.} Consider the examples shown in \cref{box:nl2sva_machine_failure_modes}: with three hand-crafted in-context examples, the same model inference under greedy decoding is able to generate functionally correct SVA implementations for the case of \texttt{gpt-4o}. However, in some cases the fixed in-context examples may serve as distractors and negatively immpact generation, especially for models with smaller parameter count such as \texttt{Llama-3.1-8B}. We posit that further improvements schemes of generating or retrieving appropriate in-context examples could therefore improve LLM capabilities in generating SystemVerilog assertions. 
\cref{tab:nl2sva_machine_result_with_sampling} further elucidate how parallel sampling of responses could realistically enable LLM-generation of functionally correct SVA implementations: we see that for \texttt{LLama-3.1-70B} a functional \texttt{pass@5} of 0.566 which greatly improves from the \texttt{pass@1} of 0.457. 

\subsection{Design2SVA Results}

Finally, Design2SVA tests if the models can produce plausible assertions directly from design RTL.
Besides the logical reasoning capabilities required in the NL2SVA-Human and NL2SVA-Machine tasks, Design2SVA further requires understanding the semantics of the given RTL implementation and also what concrete implementation of a formal property would be relevant for the design.
The synthetic RTL test cases (pipeline and FSM examples) are generated from simple more complex examples, with the most complex instances taking greater than 16K tokens when provided as context to all of the LLMs considered.
As such, we do not consider the Llama-3 family models and older OpenAI models with less than a 32K context window for this task.
\begin{table}[ht]
    \centering
\resizebox{0.9\textwidth}{!}{%
\begin{tabular}{l|cc|cc|cc|cc}
    \toprule
    \multirow{2}{*}{\textbf{Model}} & 
    \multicolumn{4}{c|}{\textbf{\texttt{Pipeline}}} & 
    \multicolumn{4}{c}{\textbf{\texttt{FSM}}}\\
           & \texttt{Syntax@1} & \texttt{Syntax@5} & \texttt{Func.@1} & \texttt{Func.@5}  
           & \texttt{Syntax@1} & \texttt{Syntax@5} & \texttt{Func.@1} & \texttt{Func.@5} \\
     \midrule
gpt-4o &  0.802 & 1.000 & 0.104 & 0.427 &  \underline{0.993} & 1.000 & 0.373 & \underline{0.900} \\
gemini-1.5-pro &  0.665 & 1.000 & \textbf{0.175} & 0.500 & 0.950 & 1.000 & \textbf{0.427} & \textbf{0.906} \\
gemini-1.5-flash & \textbf{0.969} & 1.000 & 0.025 & 0.125  & \textbf{0.996} & 1.000 & 0.079 & 0.281 \\
\midrule
Mixtral-8x22b  & 0.867 & 1.000 & 0.119 & 0.472 & 0.974 & 1.000 & 0.054 & 0.167 \\
Llama-3.1-70b & \underline{0.960} & 1.000 & \underline{0.167} & \textbf{0.615} & 0.940 & 1.000 & 0.231 & 0.719 \\
Llama-3.1-8b & 0.904 & 1.000 & 0.150 & \underline{0.552} & 0.906 & 1.000 & 0.121 & 0.521 \\
    \bottomrule
\end{tabular}%
}
\caption{Evaluation results for the Design2SVA Benchmark. Boldface and underlined results indicate highest and second highest metric values, respectively. While even the smaller models achieve moderately high syntax correctness (>80\%) with in-context examples, we see a limitations in functional correctness and again a gap between full and partial correctness of model-generated assertions.}
\label{tab:design2sva_result}
\end{table}

\begin{tcolorbox}[width=1.0\linewidth, halign=left, colframe=black, colback=white, boxsep=0.01mm, arc=1.5mm, left=2mm, right=2mm, boxrule=1pt]
\footnotesize{
\textbf{Problem ID: \texttt{fsm\_ni\_4\_nn\_4\_ne\_8\_wd\_32\_opd\_2\_0}}\\
\vspace{1em}
Question: generate 1 SVA assertion(s) for the given design RTL that is most important to verify.
If necessary, produce any extra code, including wires, registers, and their assignments.

\begin{lstlisting}[language=Verilog,  basicstyle=\BeraMonottfamily\scriptsize]
`define WIDTH 32
module fsm(clk, reset_, in_A, in_B, in_C, in_D, fsm_out);
    parameter WIDTH = `WIDTH, FSM_WIDTH = 2;
    parameter S0 = 2'b00, S1 = 2'b01, S2 = 2'b10, S3 = 2'b11;
    input clk, reset_; 
    input [WIDTH-1:0] in_A, in_B, in_C, in_D;
    output reg [FSM_WIDTH-1:0] fsm_out;
    reg [FSM_WIDTH-1:0] state, next_state;
    always_ff @(posedge clk or negedge reset_) begin
        if (!reset_) state <= S0;
        else state <= next_state;
    end
    always_comb begin
        case(state)
            S0: next_state = S2;
            S1: next_state = S3;
            S2: if ((in_D || in_C) == 'd0) next_state = S0;
                else if ((in_C <= 'd1) != in_A) next_state = S1;
                else next_state = S3;
            S3: next_state = S1;
        endcase
    end
endmodule
\end{lstlisting}
\vspace{1em}

\textbf{\texttt{gpt-4o} | Attempt 1:}\\
\begin{lstlisting}[language=Verilog,  basicstyle=\BeraMonottfamily\scriptsize]

assert property (@(posedge clk) disable iff (tb_reset)
    (state == 2'b10) |-> ##1 ((in_D == 'd0 && in_C == 'd0) || ((in_C <= 'd1) != in_A) || (next_state == 2'b11))
);
\end{lstlisting}
Syntax: \textcolor{blue}{pass} | Functionality (is proven): \textcolor{red}{fail} \\

\vspace{1em}

\textbf{\texttt{gpt-4o} | Attempt 2:}\\
\begin{lstlisting}[language=Verilog,  basicstyle=\BeraMonottfamily\scriptsize]

assert property (@(posedge clk) disable iff (tb_reset)
    state == 2'b10 |-> (next_state == 2'b00 || next_state == 2'b01 || next_state == 2'b11)
);
\end{lstlisting}
Syntax: \textcolor{blue}{pass} | Functionality  (is proven): \textcolor{blue}{pass} \\
}
\end{tcolorbox}
\noindent\begin{minipage}{\textwidth}
\captionof{figure}{Examples of failure modes observed for the Design2SVA benchmark. Multiple responses from the same language modeled are sampled, one of which incorrectly models the design RTL in formulating the assertion while the other attempt accurately identifies and implements a SystemVerilog assertion that is proven by a commercial FV tool. }\label{box:design2sva_failure_modes}
\end{minipage}

\cref{tab:design2sva_result} shows the \texttt{pass@k} values for syntax and functional correctness, as defined. 
The reported \texttt{pass@k} are based on the unbiased estimator of \texttt{pass@k} as described in \cite{Chen2021EvaluatingLL} and widely used in the literature. As we are generating multiple samples of LLM outputs, we use nucleus sampling with top $p=0.95$ and temperature $t=0.8$ as in \cite{liu2023verilogeval}.

Results show that despite most models achieve near-perfect \texttt{pass@5} for syntax correctness, the observed functional correctness---measured by whether the model has generated an assertion that can be proven---varies significantly by model.
The latest proprietary models such as \texttt{gemini-1.5-pro} and \texttt{gpt-4o} achieve considerably high \texttt{pass@5} for functional correctness for the \texttt{FSM} test instances, whereas the open-source models achieve a lower \texttt{pass@k} for most cases. Between the open-source Llama 3.1 models, we observe a significant performance gap between the larger and smaller scale models.

\cref{box:design2sva_failure_modes} highlights common scenarios where the same LLM generates multiple candidates of assertions given just the design RTL. In the case of the FSM test instance, we observe that LLMs are able to identify formal properties \emph{without any human guidance or specification} and generate correct SystemVerilog assertions that are \emph{ proven by commercial tools}. With the success rate of \texttt{pass@k} shown in \cref{tab:design2sva_result}, we anticipate LLMs to be possibly utilized to ``draft'' formal testbenches, providing a list of possible assertions to be generated for the DUT, from which the user engineer may take any assertions directly or modify as needed.
\section{Related Work}
\label{section:related_work}
\textbf{LLMs for Chip Design and Formal Verification.} 
Researchers have applied language models, and more broadly deep learning to various problems in chip design \cite{Pearce2020DAVEDA, Blocklove2023ChipChatCA, Frederiksen2020AutomatedAG, Zhao2019AutomaticAG, Kang2023LFPSLF}.
Recent work fine-tune open-source language models such as CodeGen \cite{nijkamp2022codegen} and LLaMA2 \cite{touvron2023llama} on chip design related data and show that fine-tuned models out-perform both open-source and proprietary models in Verilog code generation \cite{Thakur2023VeriGen, liu2023verilogeval} and EDA tool script generation \cite{He2023ChatEDA}. 
Prior work on applying LLMs to hardware formal verification can be broadly categorized into two groups: one line of work takes micro-architectural descriptions of the design-under-test (DUT) and prompts LLMs to generate a plausible formal testbench \cite{OrenesVera2023autosva2, sun2023nl2sva, Kande2023LLM_assertions_nyu, HassanLLMguidedFV}. 
Another line of work instead have LLMs consume architectural specification documents, describing high-level specifications that have not yet been formulated as RTL implementations, and have the models create an entire testbench \cite{Fang2024AssertLLM}. 
However, existing evaluations of LLMs on FV \cite{OrenesVera2023autosva2, Fang2024AssertLLM, liu2024domain,sun2023nl2sva} are limited to less than $\sim$20 cases of test instances and have considered a limited variety of task settings.
In contrast, \OURS expands both the scope and scale of test instances, combining collected human-written examples from industrial experts as well as verified synthetic instances generated from parameterized templates, and considers three sub-tasks each exercising varying aspects of LLM capabilities for real-world applications in FV.

\textbf{Benchmarks for Code Generation and Formal Logical Reasoning.}
Several benchmarks exist to measure language model capabilities on code generation, mostly targeting code synthesis in software programming languages (Python, C/C++, Java, etc.)
\cite{humaneval, Austin2021ProgramSW_MBPP, Chen2021EvaluatingLL}. 
Recent iterations have considered extensions that explore multi-lingual~\cite{Athiwaratkun2022MultilingualCodeEval} contexts, are geared towards particular applications such as data science~\cite{Lai2022DS1000AN}, and consider repository-level editing to reflect scenarios that are closer to real-world software engineering tasks~\cite{Zhang2024NaturalCodeBenchEC, Jimenez2023SWEbenchCL, Dinh2023LargeLMBuggyCode}.
Most related to our work, VerilogEval~\cite{liu2023verilogeval} produces a comprehensive evaluation benchmark for Verilog code generation that matches the number of test cases as software language equivalents.
Our proposed benchmark, \OURS, extends the ongoing line of research by suggesting an evaluation suite that is: (1) targeting a novel problem domain of RTL formal verification, and (2) the first to offer comprehensive evaluation for this domain, both in terms of scale of test suite size and diversity of tasks considered that mirror industrial FV workflows.

Our work also naturally bears relationships to a larger body of work on benchmarking language models on their logical reasoning capabilities, such as formal-symbolic reasoning on propositional and first-order logic~\cite{Tafjord2020ProofWriterGI, Han2022FOLIONL, Saparov2022LanguageMAprontoqa, parmar2024towards}. 
Recent work has explored neuro-symbolic approaches to enhance multi-step reasoning~\cite{Pan2023LogicLMEL, Olausson2023LINCAN, Zhou2024DontTV, Poesia2023CertifiedRW, Zelikman2022ParselAR} and have been applied to contexts of automated theorem proving~\cite{Xin2023LEGOProverNT, Yang2023LeanDojoTP}. 
LLM evaluation in formal verification can be thought of as extension to this line of work by considering modal temporal logic, where formulae include temporal operators.
\OURS is also grounded in hardware assertion code generation tasks, with natural language descriptions of the temporal logic.
Altogether, this work proposes an evaluation of LLMs on a novel class of modal logic, which can shed light on and invite future work on formal temporal reasoning, and in the context of tasks with real-world motivations and with significant value to the chip design industry.

\section{Limitations and Future Work}
The current state of \OURS is only the first step towards understanding and thereby improving model capabilities for hardware FV.
We expect future work to encompass further variety in the designs and testbenches, such as considering synthetic data generation with different styles of design modules besides the arithmetic pipeline and FSMs considered in this work.
We also anticipate methodologies that will improve LLM performance on our task, e.g. carefully chosen prompting schemes \cite{Wei2022CoT} and other structural inference methods \cite{yao2023react, Yao2023TreeOT}, as well as ideas to incorporate tool-feedback or external symbolic reasoning tools as part of a LLM-agentic framework.
Evaluations made in this work have not considered such variety of techniques and we invite future work to explore novel avenues based on \OURS.

\section{Conclusion}
In this work, we present \OURS---a first comprehensive benchmark for evaluating language models in the context of hardware formal verification.
We demonstrate three sub-tasks that are derived from industrial FV use cases and each examine different aspects of LLM capabilities necessary for their application in this domain.
\OURS defines evaluation metrics for each task with end-to-end automated evaluation made possible with a industry-standard formal verification tool. 
In particular, the custom-implemented formal equivalence checking between model generated assertions against ground-truth solution offers, for the first time, a concrete metric for evaluating LLMs in tasks generating formal properties in SVA syntax.

\label{section:conclusion}
\section*{Acknowledgements}
Authors deeply thank Vigyan Singhal for technical guidance and support. 
We also thank Cadence Design Systems for the implementation of the formal assertion-to-assertion equivalence checking in Jasper and contributions to the overall project, including helpful discussions.

\bibliography{fveval}
\bibliographystyle{plain}
\newpage
\appendix
\newpage
\section{Additional Details on \OURS: NL2SVA-Human}
In this section, we provide additional details and test instance examples for NL2SVA-Human.

\subsection{NLSVA-Human Benchmark Statistics and Examples}
\label{subsec:appendix_nl2sva_human_examples}

Detailed statistics on the composition of formal testbenches consisting of the NL2SVA-Human benchmark are shown in Table~\ref{tab:nl2sva-human-stats}.
\begin{table}[ht]
\centering
\resizebox{0.5\textwidth}{!}{%
\begin{tabular}{|l|c|c|} 
  \hline
 Name & \# Variations & \# Assertions  \\
 \hline
 \hline
  1R1W FIFO & 4  & 20 \\
  Multi-Port FIFO & 1  & 6 \\
 Arbiter & 4 & 37 \\
 FSM & 2 & 4  \\
 Counter & 1 & 5\\
 RAM & 1 & 7 \\
 \hline
\hline
Total & 13 & 79 \\
 \hline
\end{tabular}
}
\vspace{10pt}
\caption{Statistics describing the NL2SVA-Human benchmark. From industrial formal testbenches, we extract a total of 79 test cases of NL specification to SVA assertion.}
\label{tab:nl2sva-human-stats}
\end{table}

Next, we show an example testbench in NL2SVA-Human and its associated assertions and NL specifications:
\newpage
\begin{tcolorbox}[width=1.0\linewidth, halign=left, colframe=black, colback=white, boxsep=0.01mm, arc=1.5mm, left=2mm, right=2mm, boxrule=1pt]
\footnotesize{
\textbf{Textbench Example: 1R1W FIFO}}
\end{tcolorbox}

\begin{lstlisting}[language=Verilog]
module fifo_1r1w_tb (
     clk,
     reset_,
     wr_vld,
     wr_data,
     wr_ready,
     rd_vld,
     rd_data,
     rd_ready
   );
  
    parameter   FIFO_DEPTH                              = 4;
    parameter   DATA_WIDTH                              = 1;

    localparam FIFO_DEPTH_log2 = $clog2(FIFO_DEPTH); 
    localparam DATA_WIDTH_log2 = $clog2(DATA_WIDTH); 

    input                   clk;
    input                   reset_;
    input                   wr_vld;
    input  [DATA_WIDTH-1:0] wr_data;
    input                   wr_ready;
    input                   rd_vld;
    input  [DATA_WIDTH-1:0] rd_data;
    input                   rd_ready;

wire wr_push;
wire rd_pop;

wire tb_reset;
assign tb_reset = (reset_ == 1'b0);

wire fifo_full;
assign wr_push = wr_vld && wr_ready;
assign rd_pop = rd_vld && rd_ready; 

//fifo array - shift register
reg [DATA_WIDTH-1:0]             fifo_array [FIFO_DEPTH-1:0]; 
//fifo array - rd_ptr
reg [FIFO_DEPTH_log2-1:0]        fifo_rd_ptr;    
// actual pop == pop
wire                             actual_fifo_pop;       
// fifo empty
reg                              fifo_empty;   
// dout
wire [DATA_WIDTH-1:0]            fifo_out_data;               

// ---- shift register code start ----
always @(posedge clk) begin
    if (!reset_) fifo_array[0] <= 'd0;
    else if (wr_push) begin
        fifo_array[0] <= wr_data;
    end else fifo_array[0] <= fifo_array[0];
end
for (genvar i = 1; i < FIFO_DEPTH; i++ ) begin : loop_id
    always @(posedge clk) begin
        if (!reset_) fifo_array[i] <= 'd0;
        else if (wr_push) begin
            fifo_array[i] <= fifo_array[i-1];
        end else fifo_array[i] <= fifo_array[i];
    end
end

// ---- read pointer/fifo empty code start ----
always @(posedge clk) begin
    if (!reset_) begin
        fifo_rd_ptr <= 'd0;
    end else if (wr_push && fifo_empty)  begin
        fifo_rd_ptr <= 'd0;
    end else if (rd_pop && !fifo_empty && (fifo_rd_ptr == 'd0)) begin
        fifo_rd_ptr <= 'd0;
    end else begin
        fifo_rd_ptr <= fifo_rd_ptr + wr_push - rd_pop;
    end
    if (!reset_) begin
        fifo_empty <= 'd1;
    end else if (rd_pop && !fifo_empty && (fifo_rd_ptr == 'd0) && !wr_push) begin
        fifo_empty <= 'd1;
    end else if ((fifo_rd_ptr != 'd0) || wr_push && !rd_pop) begin
        fifo_empty <= 'd0;
    end
end
// ---- fifo full and dout code start ----
assign fifo_full = (fifo_rd_ptr == (FIFO_DEPTH - 1)) && !fifo_empty;
assign fifo_out_data = fifo_array[fifo_rd_ptr];

endmodule
\end{lstlisting}
\noindent\begin{minipage}{\textwidth}
\captionof{figure}{An example testbench instance (one version of the 1R1W FIFO) included in the NL2SVA-Human benchmark.
Each testbench RTL (in SystemVerilog as shown) is provided as context to the LLMs during evaluation.
For the full prompt format we use, see Figure~\ref{box:nl2sva_human_prompt}.
}\label{box:nl2sva_human_testbench_example}
\end{minipage}
\newpage

\begin{tcolorbox}[width=1.0\linewidth, halign=left, colframe=black, colback=white, boxsep=0.01mm, arc=1.5mm, left=2mm, right=2mm, boxrule=1pt]
\footnotesize{
\textbf{NL Specification and SVA Assertion Pairs for 1R1W FIFO}

\vspace{5pt}
Question: Create a SVA assertion that checks: that the FIFO does not underflow, assuming no bypass. Use the signals 'rd\_pop' and 'fifo\_empty'.
\vspace{5pt}

Answer:
\begin{lstlisting}[language=Verilog]
asrt : assert property (@(posedge clk) disable iff (tb_reset)
    (fifo_empty && rd_pop) !== 1'b1
);
\end{lstlisting}

Question: Create a SVA assertion that checks: that the FIFO does not overflow, assuming no bypass. Use the signals 'wr\_push' and 'fifo\_full'.
\vspace{5pt}

Answer:
\begin{lstlisting}[language=Verilog]
asrt : assert property (@(posedge clk) disable iff (tb_reset)
    (fifo_full && wr_push) !== 1'b1
);
\end{lstlisting}

Question: Create a SVA assertion that checks: that the fifo output and read data are consistent, assuming no bypass. Use the signals 'rd\_pop', 'rd\_data', and 'fifo\_out\_data'.
\vspace{5pt}

Answer:
\begin{lstlisting}[language=Verilog]
asrt: assert property (@(posedge clk) disable iff (tb_reset)
       (rd_pop && (fifo_out_data != rd_data)) !== 1'b1
);
\end{lstlisting}

Question: Create a SVA assertion that checks: that when response is pending, data is eventually popped from the FIFO. Use the signals 'rd\_pop' and 'fifo\_empty'.
\vspace{5pt}

Answer:
\begin{lstlisting}[language=Verilog]
asrt: assert property (@(posedge clk) disable iff (tb_reset)
       !fifo_empty |-> strong(##[0:$] rd_pop)
);
\end{lstlisting}

Question: Create a SVA assertion that checks: that when there is a write push to the FIFO, data is eventually popped. Use the signals 'rd\_pop' and 'wr\_push'.
\vspace{5pt}

Answer:
\begin{lstlisting}[language=Verilog]
asrt: assert property (@(posedge clk) disable iff (tb_reset)
       wr_push |-> strong(##[0:$] rd_pop)
);
\end{lstlisting}
}
\end{tcolorbox}
\noindent\begin{minipage}{\textwidth}
\captionof{figure}{List of natural language (NL) specification to SVA assertion implementations that are collected for the example 1R1W FIFO test bench shown above in ~\ref{box:nl2sva_human_testbench_example}.
There are a total of five test instances for this testbench.
}\label{box:nl2sva_human_testbench_example_questions}
\end{minipage}

\newpage
\subsection{NL2SVA-Human Evaluation: Model Prompt}
\label{subsec:appendix_nl2sva_human_prompt}
The following figure describes the prompt provided to language models during evaluation.
\begin{tcolorbox}[width=1.0\linewidth, halign=left, colframe=black, colback=white, boxsep=0.01mm, arc=1.5mm, left=2mm, right=2mm, boxrule=1pt]
\footnotesize{
\textbf{System:}
\vspace{5pt}

You are an AI assistant tasked with formal verification of register transfer level (RTL) designs. 

Your job is to translate a description of an assertion to concrete SystemVerilog Assertion (SVA) implementation. 

}

\vspace{5pt}
\textbf{User:}
\vspace{5pt}

Here is the testbench to perform your translation:
\begin{lstlisting}[language=Verilog]
[Testbench RTL (fifo_1r1w.sv)]
\end{lstlisting}

Question: Create a SVA assertion that checks: 
that the fifo output and read data are consistent, assuming no bypass. Use the signals 'rd\_pop', 'rd\_data', and 'fifo\_out\_data'.

\vspace{5pt}
Do not add code to output an error message string.
Enclose your SVA code with ```systemverilog and ```. 
Only output the code snippet and do NOT output anything else.

For example,
\begin{lstlisting}[language=Verilog]
```systemverilog
asrt: assert property (@(posedge clk) disable iff (tb_reset)
    (a && b) != 1'b1
);
```
\end{lstlisting}

Answer:
\end{tcolorbox}
\noindent\begin{minipage}{\textwidth}
\captionof{figure}{Details of the prompt given to LLMs for the NL2SVA-Human benchmark.
}\label{box:nl2sva_human_prompt}
\end{minipage}

\newpage
\subsection{NL2SVA-Human Example Model Responses}
Here, we show sample LLM responses to the example test instance (testbench and NL specification of assertion) referred to in Section~\ref{subsec:appendix_nl2sva_human_examples} and ~\ref{subsec:appendix_nl2sva_human_prompt}.
\label{subsec:appendix_nl2sva_human_model_responses}

\begin{tcolorbox}[width=1.0\linewidth, halign=left, colframe=black, colback=white, boxsep=0.01mm, arc=1.5mm, left=2mm, right=2mm, boxrule=1pt]
\footnotesize{
\textbf{Examples of LLM Responses to NL2SVA-Human (1R1W FIFO)}
\vspace{5pt}

\textbf{Question:} Create a SVA assertion that checks: that the fifo output and read data are consistent, assuming no bypass. Use the signals 'rd\_pop', 'rd\_data', and 'fifo\_out\_data'.
\vspace{5pt}

\textbf{Reference Solution:}
\begin{lstlisting}[language=Verilog]
asrt: assert property (@(posedge clk) disable iff (tb_reset)
       (rd_pop && (fifo_out_data != rd_data)) !== 1'b1
);
\end{lstlisting}

\textbf{gpt-4-0613:}
\begin{lstlisting}[language=Verilog]
asrt_fifo_out_rd_data_consistency: assert property (@(posedge clk) disable iff (tb_reset)
    rd_pop |-> (rd_data == fifo_out_data)
);
\end{lstlisting}

\textbf{gpt-4o-05-13:}
\begin{lstlisting}[language=Verilog]
asrt_fifo_output_consistency: assert property (@(posedge clk) disable iff (tb_reset)
    rd_pop |-> (rd_data == fifo_out_data)
);
\end{lstlisting}
\textbf{gpt-3.5-turbo-0125:}
\begin{lstlisting}[language=Verilog]
asrt: assert property (@(posedge clk)
    disable iff (tb_reset)
    (rd_pop && (rd_data !== fifo_out_data)) | (!rd_pop && (rd_data === fifo_out_data))
);
\end{lstlisting}
\textbf{Llama3-70B-chat:}
\begin{lstlisting}[language=Verilog]
asrt_fifo_out_data: assert property (@(posedge clk) disable iff (tb_reset)
    rd_pop |-> (rd_data == fifo_out_data)
);
\end{lstlisting}
\textbf{Mixtral-8x22B-instruct-v1:}
\begin{lstlisting}[language=Verilog]
asrt_fifo_output_consistency: assert property (@(posedge clk) disable iff (tb_reset)
    rd_pop |-> rd_data === fifo_out_data
);
\end{lstlisting}
\textbf{Llama3-8B-chat:}
\begin{lstlisting}[language=Verilog]
asrt: assert property (@(posedge clk) disable iff (tb_reset)
    rd_pop |-> $rose(fifo_rd_ptr) |=> rd_data == fifo_out_data
);
\end{lstlisting}
\textbf{Mixtral-8x7B-instruct-v1:}
\begin{lstlisting}[language=Verilog]
asrt: assert property (@(posedge clk) disable iff (tb_reset)
    !((rd_pop && rd_data !== fifo_out_data) && !fifo_empty)
);
\end{lstlisting}

}
\end{tcolorbox}
\noindent\begin{minipage}{\textwidth}
\captionof{figure}{Example responses from different LLMs for the NL2SVA-Human test instance shown in Figure ~\ref{box:nl2sva_human_prompt}.}
\label{box:nl2sva_human_model_response}
\end{minipage}
\newpage

\section{Additional Details on \OURS: NL2SVA-Machine}
In this section, we provide additional details and test instance examples for NL2SVA-Machine.

\subsection{NL2SVA-Machine Evaluation: Model Prompt}
\label{subsec:appendix_nl2sva_machine_prompt}
The following figure describes the prompt provided to language models during evaluation.
The prompt shown is for zero-shot evaluation; for $k=3$-shot, we include the ICL examples listed in Figure~\ref{box:nl2sva_machine_icl_prompt} as part of the user prompt, before the indicated question.
\begin{tcolorbox}[width=1.0\linewidth, halign=left, colframe=black, colback=white, boxsep=0.01mm, arc=1.5mm, left=2mm, right=2mm, boxrule=1pt]
\footnotesize{
\textbf{System:}
\vspace{5pt}

You are an AI assistant tasked with formal verification of register transfer level (RTL) designs. 

Your job is to translate a description of an assertion to concrete SystemVerilog Assertion (SVA) implementation.
}

\vspace{5pt}
\textbf{User:}
\vspace{5pt}

Question: Create a SVA assertion that checks: 

If sig\_G is not equal to 1'b1, then four cycles later, sig\_J must be true.
\vspace{5pt}

Do not add code to output an error message string.
Enclose your SVA code with ```systemverilog and ```. 
Only output the code snippet and do NOT output anything else.

For example,
\begin{lstlisting}[language=Verilog]
```systemverilog
asrt: assert property (@(posedge clk) disable iff (tb_reset)
    (sig_A && sig_B) != 1'b1
);
```
\end{lstlisting}

Answer:
\end{tcolorbox}
\noindent\begin{minipage}{\textwidth}
\captionof{figure}{Details of the prompt given to LLMs for the NL2SVA-Machine benchmark.
When models are evaluated in 3-shot, we add the fixed ICL examples shown in Figure~\ref{box:nl2sva_machine_icl_prompt}.
}\label{box:nl2sva_machine_prompt}
\end{minipage}
\newpage

\subsection{NL2SVA-Machine Evaluation: In-Context Examples Used}
\label{subsec:appendix_nl2sva_machine_prompt_icl}
The following lists the in-context examples used in NL2SVA-Machine:
\begin{tcolorbox}[width=1.0\linewidth, halign=left, colframe=black, colback=white, boxsep=0.01mm, arc=1.5mm, left=2mm, right=2mm, boxrule=1pt]
\footnotesize{
\textbf{ICL Examples}
\vspace{5pt}

More detailed examples of correct translations from description into an SVA assertion:

\vspace{5pt}

Question: Create a SVA assertion that checks: Whenever sig\_A is high and sig\_B is low, sig\_C will be high on the next clock edge.
Do not add code to output an error message string.
Enclose your SVA code with ```systemverilog and ```. Only output the code snippet and do NOT output anything else.

Answer:
\begin{lstlisting}[language=Verilog]
```systemverilog
assert property(@(posedge clk)
    (sig_A && !sig_B) |-> sig_C
);
```
\end{lstlisting}

Question: Create a SVA assertion that checks: If sig\_C contains at least one '1' bit or sig\_D is not equal to sig\_A, then sig\_F must eventually be true
Do not add code to output an error message string.
Enclose your SVA code with ```systemverilog and ```. Only output the code snippet and do NOT output anything else.

Answer:
\begin{lstlisting}[language=Verilog]
```systemverilog
assert property(@(posedge clk)
    (|sig_C || (sig_D !== sig_A )) |=> s_eventually(sig_F)
);
```
\end{lstlisting}

Question: Create a SVA assertion that checks: "Whenever the value of sig\_J is less than the result of the XOR operation between sig\_C and the negation of the bitwise negation of sig\_H, and this result is equal to the result of the OR operation between the identity comparison of sig\_A and the negation of sig\_J and sig\_B, the assertion is true
Do not add code to output an error message string.
Enclose your SVA code with ```systemverilog and ```. Only output the code snippet and do NOT output anything else.

Answer:
\begin{lstlisting}[language=Verilog]
```systemverilog
assert property(@(posedge clk)
	((sig_J < (sig_B == (sig_C ^ ~|sig_H))) == ((|sig_A === !sig_J) || sig_B))
);
```
\end{lstlisting}
}
\end{tcolorbox}
\noindent\begin{minipage}{\textwidth}
\captionof{figure}[hypcap=false]{Details of the ICL examples, given as part of the prompt to LLMs for the NL2SVA-Machine benchmark for 3-shot performance evaluation. Note that the same examples are used across all test cases in our study.}\label{box:nl2sva_machine_icl_prompt}
\end{minipage}
\newpage

\subsection{NL2SVA-Machine Example Model Responses}
\label{subsec:appendix_nl2sva_machine_model_responses}
Here, we show sample LLM responses to the example test instance referred to in Section~\ref{subsec:appendix_nl2sva_machine_prompt}.
\begin{tcolorbox}[width=1.0\linewidth, halign=left, colframe=black, colback=white, boxsep=0.01mm, arc=1.5mm, left=2mm, right=2mm, boxrule=1pt]
\footnotesize{
\textbf{Examples of LLM Responses to NL2SVA-Human (1R1W FIFO)}
\vspace{5pt}

\textbf{Question:} Create a SVA assertion that checks: If sig\_G is not equal to 1'b1, then four cycles later, sig\_J must be true.
\vspace{5pt}

\textbf{Reference Solution:}
\begin{lstlisting}[language=Verilog]
assert property(@(posedge clk)
    (sig_G !== 1'b1) |-> ##4 sig_J
);
\end{lstlisting}

\textbf{gpt-4-0613:}
\begin{lstlisting}[language=Verilog]
assert property(@(posedge clk)
    (sig_G !== 1'b1) |-> ##4 sig_J
);
\end{lstlisting}

\textbf{gpt-4o-05-13:}
\begin{lstlisting}[language=Verilog]
assert property(@(posedge clk)
    (sig_G !== 1'b1) |-> ##4 sig_J
);
\end{lstlisting}
\textbf{gpt-3.5-turbo-0125:}
\begin{lstlisting}[language=Verilog]
assert property(@(posedge clk)
    (sig_G !== 1'b1) |-> ##4 sig_J
);
\end{lstlisting}
\textbf{Llama3-70B-chat:}
\begin{lstlisting}[language=Verilog]
assert property(@(posedge clk)
    (!sig_G) |-> ##[4] sig_J
);
\end{lstlisting}
\textbf{Mixtral-8x22B-instruct-v1:}
\begin{lstlisting}[language=Verilog]
assert property(@(posedge clk)
    sig_G !== 1'b1 |-> ##4 sig_J
);
\end{lstlisting}
\textbf{Llama3-8B-chat:}
\begin{lstlisting}[language=Verilog]
assert property(@(posedge clk) ($rose(!sig_G) |=> ##[3] sig_J));
\end{lstlisting}
\textbf{Mixtral-8x7B-instruct-v1:}
\begin{lstlisting}[language=Verilog]
assert property(@(posedge clk)
    (sig_G !== 1'b1) |-> ##[1:4] sig_J
);
\end{lstlisting}
}
\end{tcolorbox}
\noindent\begin{minipage}{\textwidth}
\captionof{figure}{Example responses from different LLMs for the NL2SVA-Machine test instance shown in Figure~\ref{box:nl2sva_machine_prompt}.
The OpenAI models (\texttt{gpt-4}, \texttt{gpt-4o}, and \texttt{gpt-3.5-turbo}) and the larger Mixtral MoE model generates SVA assertions that exactly matches the reference solution.
}\label{box:nl2sva_machine_model_response}
\end{minipage}

\newpage
\section{Additional Details on \OURS: Design2SVA}
In this section, we provide additional details and test instance examples for Design2SVA.

\subsection{Examples of Design2SVA RTL Test Cases}
The following demonstrates an example design RTL generated for the Design2SVA benchmark.
As noted, we generate synthetic test cases representing arithmetic pipelines and finite state machines (FSMs).
Here we show the simplest examples from each category.
Each generated design is accompanied with a testbench harness, of which the module definition and input/output port information is provided as context to LLMs, as described in Section~\ref{subsec:appendix_deisng2sva_prompt}.
\label{subsec:appendix_deisng2sva_examples}
Synthetic pipeline design RTL and testbench header example:
\begin{tcolorbox}[width=1.0\linewidth, halign=left, colframe=black, colback=white, boxsep=0.01mm, arc=1.5mm, left=2mm, right=2mm, boxrule=1pt]
\footnotesize{
\textbf{Design2SVA Design RTL Example (Pipeline)}}
\end{tcolorbox}

\begin{lstlisting}[language=Verilog]
`define WIDTH 128
`define DEPTH 6

module exec_unit_0 (
    clk,
    reset_,
    in_data,
    in_vld,
    out_data,
    out_vld
);  
    parameter WIDTH = `WIDTH;
    localparam DEPTH = 4;
    input clk;
    input reset_;
    input [WIDTH-1:0] in_data;
    input in_vld;
    output [WIDTH-1:0] out_data;
    output out_vld;

    
    logic [DEPTH:0] ready;
    logic [DEPTH:0][WIDTH-1:0] data;
    assign ready[0] = in_vld;
    assign data[0] = in_data;
    assign out_vld = ready[DEPTH];
    assign out_data = data[DEPTH];

    generate
        for (genvar i=0; i < DEPTH; i=i+1) begin : gen
            always @(posedge clk) begin
                if (!reset_) begin
                    ready[i+1] <= 'd0;
                    data[i+1] <= 'd0;
                end else begin
                    ready[i+1] <= ready[i];
                    data[i+1] <= ((((data[i] ^ 9) - ((data[i] <<< 7) + (data[i] - 5))) >>> 8) + 4);
                end
            end
        end
    endgenerate
endmodule


module exec_unit_1 (
    clk,
    reset_,
    in_data,
    in_vld,
    out_data,
    out_vld
);  
    parameter WIDTH = `WIDTH;
    localparam DEPTH = 2;
    input clk;
    input reset_;
    input [WIDTH-1:0] in_data;
    input in_vld;
    output [WIDTH-1:0] out_data;
    output out_vld;

    
    logic [DEPTH:0] ready;
    logic [DEPTH:0][WIDTH-1:0] data;
    assign ready[0] = in_vld;
    assign data[0] = in_data;
    assign out_vld = ready[DEPTH];
    assign out_data = data[DEPTH];

    generate
        for (genvar i=0; i < DEPTH; i=i+1) begin : gen
            always @(posedge clk) begin
                if (!reset_) begin
                    ready[i+1] <= 'd0;
                    data[i+1] <= 'd0;
                end else begin
                    ready[i+1] <= ready[i];
                    data[i+1] <= ((((data[i] - 5) >>> 1) >>> 7) ^ 7);
                end
            end
        end
    endgenerate
endmodule



module pipeline (
    clk,
    reset_,
    in_vld,
    in_data,
    out_vld,
    out_data
);
    parameter WIDTH=`WIDTH;
    parameter DEPTH=`DEPTH;
    
    input clk;
    input reset_;
    input in_vld;
    input [WIDTH-1:0] in_data;
    output out_vld;
    output [WIDTH-1:0] out_data;

    wire [DEPTH:0] ready;
    wire [DEPTH:0][WIDTH-1:0] data;
    assign ready[0] = in_vld;
    assign data[0] = in_data;
    assign out_vld = ready[DEPTH];
    assign out_data = data[DEPTH];

    exec_unit_0 #(.WIDTH(WIDTH)) unit_0 (
        .clk(clk),
        .reset_(reset_),
        .in_data(data[0]),
        .in_vld(ready[0]),
        .out_data(data[4]), 
        .out_vld(ready[4])
    );

    exec_unit_1 #(.WIDTH(WIDTH)) unit_1 (
        .clk(clk),
        .reset_(reset_),
        .in_data(data[4]),
        .in_vld(ready[4]),
        .out_data(data[6]), 
        .out_vld(ready[6])
    );
endmodule
\end{lstlisting}
\noindent\begin{minipage}{\textwidth}
\captionof{figure}{An example Pipeline test instance generated for Design2SVA. }\label{box:design2sva_pipeline_example}
\end{minipage}

\newpage
\begin{tcolorbox}[width=1.0\linewidth, halign=left, colframe=black, colback=white, boxsep=0.01mm, arc=1.5mm, left=2mm, right=2mm, boxrule=1pt]
\footnotesize{
\textbf{Design2SVA Testbench RTL Module Header Example (Pipeline)}}
\end{tcolorbox}

\begin{lstlisting}[language=Verilog]
`define WIDTH 128
`define DEPTH 6

module pipeline_tb (
    clk,
    reset_,
    in_vld,
    in_data,
    out_vld,
    out_data
);
    parameter WIDTH=`WIDTH;
    parameter DEPTH=`DEPTH;

    input clk;
    input reset_;
    input in_vld;
    input [WIDTH-1:0] in_data;
    input out_vld;
    input [WIDTH-1:0] out_data;

    assign tb_reset = (reset_ == 1'b0);
endmodule
\end{lstlisting}
\noindent\begin{minipage}{\textwidth}
\captionof{figure}{Module definition and header of the formal testbench accompanying the Pipeline example above for Design2SVA. }\label{box:design2sva_pipeline_example_tb}
\end{minipage}
\newpage
Synthetic FSM design RTL and testbench header example:
\begin{tcolorbox}[width=1.0\linewidth, halign=left, colframe=black, colback=white, boxsep=0.01mm, arc=1.5mm, left=2mm, right=2mm, boxrule=1pt]
\footnotesize{
\textbf{Design2SVA Design RTL Example (FSM)}}
\end{tcolorbox}

\begin{lstlisting}[language=Verilog]
`define WIDTH 32
module fsm(
    clk,
    reset_,
    in_A,
    in_B,
    in_C,
    in_D,
    fsm_out
);
    parameter WIDTH = `WIDTH;
    parameter FSM_WIDTH = 2;

    parameter S0 = 2'b00;
    parameter S1 = 2'b01;
    parameter S2 = 2'b10;
    parameter S3 = 2'b11;

    input clk;
    input reset_;
    input [WIDTH-1:0] in_A;
    input [WIDTH-1:0] in_B;
    input [WIDTH-1:0] in_C;
    input [WIDTH-1:0] in_D;
    output reg [FSM_WIDTH-1:0] fsm_out;
    reg [FSM_WIDTH-1:0] state, next_state;
    always_ff @(posedge clk or negedge reset_) begin
        if (!reset_) begin
            state <= S0;
        end else begin
            state <= next_state;
        end
    end
    always_comb begin
        case(state)
            S0: begin
                next_state = S2;
            end
            S1: begin
                next_state = S3;
            end
            S2: begin
                if (((in_A != in_B) < 'd0)) begin
                    next_state = S0;
                end
                else begin
                    next_state = S1;
                end
            end
            S3: begin
            end
        endcase
    end
endmodule
\end{lstlisting}
\noindent\begin{minipage}{\textwidth}
\captionof{figure}{An example FSM test instance generated for Design2SVA. }\label{box:design2sva_fsm_example}
\end{minipage}

\begin{tcolorbox}[width=1.0\linewidth, halign=left, colframe=black, colback=white, boxsep=0.01mm, arc=1.5mm, left=2mm, right=2mm, boxrule=1pt]
\footnotesize{
\textbf{Design2SVA Testbench RTL Module Header Example (FSM)}}
\end{tcolorbox}

\begin{lstlisting}[language=Verilog]
`define WIDTH 32
module fsm_tb(
    clk,
    reset_,
    in_A,
    in_B,
    in_C,
    in_D,
    fsm_out
);
    parameter WIDTH = `WIDTH;
    parameter FSM_WIDTH = 2;

    parameter S0 = 2'b00;
    parameter S1 = 2'b01;
    parameter S2 = 2'b10;
    parameter S3 = 2'b11;

    input clk;
    input reset_;
    input [WIDTH-1:0] in_A;
    input [WIDTH-1:0] in_B;
    input [WIDTH-1:0] in_C;
    input [WIDTH-1:0] in_D;
    input reg [FSM_WIDTH-1:0] fsm_out;
    wire tb_reset;
    assign tb_reset = (reset_ == 1'b0);
endmodule
\end{lstlisting}
\noindent\begin{minipage}{\textwidth}
\captionof{figure}{Module definition and header of the formal testbench accompanying the FSM example above for Design2SVA. }\label{box:design2sva_fsm_example_tb}
\end{minipage}
\newpage

\subsection{Design2SVA Evaluation: Model Prompt}
The following details the prompt provided to language models during evaluation of Design2SVA.
\label{subsec:appendix_deisng2sva_prompt}

\begin{tcolorbox}[width=1.0\linewidth, halign=left, colframe=black, colback=white, boxsep=0.01mm, arc=1.5mm, left=2mm, right=2mm, boxrule=1pt]
\footnotesize{
\textbf{System:}
\vspace{5pt}

You are an AI assistant tasked with formal verification of register transfer level (RTL) designs.

Your job is to generate a SystemVerilog assertion for the design-under-test provided.
}

\vspace{5pt}
\textbf{User:}
\vspace{5pt}

Here is the design RTL to generate assertions for:
\begin{lstlisting}[language=Verilog]
[Design RTL (fsm.sv)]
\end{lstlisting}

Here is a partial testbench for you to work on:
\begin{lstlisting}[language=Verilog]
[Testbench RTL Header(fsm_tb.sv)]
\end{lstlisting}

Question: generate a single SVA assertion for the given design RTL that is most important to verify.

If necessary, produce any extra code, including wires, registers, and their assignments.

Do NOT use signals from the design RTL, only use the module input signals or internal signals you have added.

Do NOT use any 'initial' blocks. This testbench is not for running RTL simulation but for formal verification.

Do NOT instantiate the design module inside the testbench.

When implementing the assertion, generate a concurrent SVA assertion and do not add code to output an error message string.

Enclose your SystemVerilog code with ```systemverilog and ```. 

For example, 
\begin{lstlisting}[language=Verilog]
```systemverilog
logic [3:0] a, b;
assign a = c & d;
assign b = c | d;

assert property (@(posedge clk) disable iff (tb_reset)
    (a && b) != 1'b1
);
```
\end{lstlisting}

Only output the code snippet and do NOT output anything else.

Remember to output only one assertion.

Answer:
\end{tcolorbox}
\noindent\begin{minipage}{\textwidth}
\captionof{figure}{Details of the prompt given to LLMs for the Design2SVA benchmark.}\label{box:design2sva_prompt}
\end{minipage}
\newpage

\subsection{Design2SVA Example Model Responses}
Here, we show sample LLM responses to the example test instance referred to in Section~\ref{subsec:appendix_deisng2sva_prompt}.
Again, we list examples for both the pipeline RTL example and the FSM RTL example.

\label{subsec:appendix_deisng2sva_model_responses}
\begin{tcolorbox}[width=1.0\linewidth, halign=left, colframe=black, colback=white, boxsep=0.01mm, arc=1.5mm, left=2mm, right=2mm, boxrule=1pt]
\footnotesize{
\textbf{Examples of LLM Responses to Design2SVA (Pipeline)}

\vspace{5pt}

{\normalsize \textbf{gpt-4-turbo (gpt-4-0125-preview):}}

\vspace{5pt}

\textbf{Attempt 1}
\begin{lstlisting}[language=Verilog]
logic [WIDTH-1:0] tb_in_data;
logic tb_in_vld, tb_out_vld;
logic [WIDTH-1:0] tb_out_data;
always @(posedge clk or negedge reset_) begin
    if (!reset_) begin
        tb_in_data <= '0;
        tb_in_vld <= 1'b0;
    end else begin
        tb_in_data <= $random;
        tb_in_vld <= $random;
    end
end

assign tb_out_vld = out_vld;
assign tb_out_data = out_data;

assert property (@(posedge clk) disable iff (!reset_)
    tb_in_vld |-> ##DEPTH tb_out_vld
);
\end{lstlisting}

\textbf{Attempt 2}
\begin{lstlisting}[language=Verilog]
logic [WIDTH-1:0] prev_data;
logic prev_data_valid;
always_ff @(posedge clk or negedge reset_) begin
    if (!reset_) begin
        prev_data <= 'd0;
        prev_data_valid <= 1'b0;
    end else if (in_vld) begin
        prev_data <= in_data;
        prev_data_valid <= 1'b1;
    end
end

assert property (@(posedge clk) disable iff (!reset_)
    prev_data_valid && out_vld |-> ##[1:DEPTH] (out_data !== 'd0)
);
\end{lstlisting}
}
\end{tcolorbox}
\noindent\begin{minipage}{\textwidth}
\captionof{figure}{Example responses from \texttt{gpt-4-turbo} for the Pipeline test instance shown in ~\ref{subsec:appendix_deisng2sva_examples}. 
Two attempts from the model are shown.}\label{box:design2sva_model_response_pipeline_gpt-4-turbo}
\end{minipage}
\newpage

\begin{tcolorbox}[width=1.0\linewidth, halign=left, colframe=black, colback=white, boxsep=0.01mm, arc=1.5mm, left=2mm, right=2mm, boxrule=1pt]
\footnotesize{
\textbf{[Continued] Examples of LLM Responses to Design2SVA (Pipeline)}

\vspace{5pt}

{\normalsize \textbf{gpt-4o-05-13:}}

\vspace{5pt}

\textbf{Attempt 1}
\begin{lstlisting}[language=Verilog]
logic [WIDTH-1:0] tb_in_data, tb_out_data;
logic tb_in_vld, tb_out_vld;
assign tb_in_data = in_data;
assign tb_out_data = out_data;
assign tb_in_vld = in_vld;
assign tb_out_vld = out_vld;

assert property (@(posedge clk) disable iff (tb_reset)
    tb_in_vld |-> 
        (tb_out_vld && tb_out_data == 
            ((((((((tb_in_data & 8) ^ (tb_in_data - 9)) & 8) ^ 
                (((tb_in_data & 8) ^ (tb_in_data - 9)) - 9)) & 8) ^ 
                    (((tb_in_data & 8) ^ (tb_in_data - 9)) - 9)) <<< 2) ^ 10) - 2)
);
\end{lstlisting}

\textbf{Attempt 2}
\begin{lstlisting}[language=Verilog]
logic [WIDTH-1:0] tb_in_data;
logic tb_in_vld, tb_out_vld, tb_reset;
logic [WIDTH-1:0] tb_out_data;

assign tb_in_data = in_data;
assign tb_in_vld = in_vld;
assign tb_out_vld = out_vld;
assign tb_out_data = out_data;

assert property (@(posedge clk) disable iff (tb_reset)
    tb_in_vld |-> 
        (tb_out_vld && tb_out_data == 
            (((((tb_in_data & 8) ^ (tb_in_data - 9)) <<< 2) ^ 10) - 2))
);
\end{lstlisting}
}
\end{tcolorbox}
\noindent\begin{minipage}{\textwidth}
\captionof{figure}{Example responses from \texttt{gpt-4o} for the Pipeline test instance shown in ~\ref{subsec:appendix_deisng2sva_examples}. 
Two attempts from the model are shown.}\label{box:design2sva_model_response_pipeline_gpt4o}
\end{minipage}

\newpage

\begin{tcolorbox}[width=1.0\linewidth, halign=left, colframe=black, colback=white, boxsep=0.01mm, arc=1.5mm, left=2mm, right=2mm, boxrule=1pt]
\footnotesize{
\textbf{[Continued] Examples of LLM Responses to Design2SVA (Pipeline)}

\vspace{5pt}

{\normalsize \textbf{Mixtral-8x22B-instruct-v1:}}

\vspace{5pt}

\textbf{Attempt 1}
\begin{lstlisting}[language=Verilog]
logic [WIDTH-1:0] data_in;
logic data_in_vld;

assign data_in = in_data;
assign data_in_vld = in_vld;

assert property (@(posedge clk) disable iff (tb_reset)
    $rose(data_in_vld) |=> ##[1:DEPTH] out_vld
);
\end{lstlisting}

\textbf{Attempt 2}
\begin{lstlisting}[language=Verilog]
logic [DEPTH-1:0] tb_in_vld;
logic [DEPTH-1:0][WIDTH-1:0] tb_in_data;
assign tb_in_vld[0] = in_vld;
assign tb_in_data[0] = in_data;

assert property (@(posedge clk) disable iff (tb_reset)
    $rose(out_vld) |-> tb_in_vld[DEPTH]
);
\end{lstlisting}
}
\end{tcolorbox}
\noindent\begin{minipage}{\textwidth}
\captionof{figure}{Example responses from \texttt{Mixtral-8x22B} for the Pipeline test instance shown in ~\ref{subsec:appendix_deisng2sva_examples}. 
Two attempts from the model are shown.}\label{box:design2sva_model_response_pipeline_mixtral_8x22b}
\end{minipage}

\newpage
Example LLM responses for FSM RTL test cases in Design2SVA:
\begin{tcolorbox}[width=1.0\linewidth, halign=left, colframe=black, colback=white, boxsep=0.01mm, arc=1.5mm, left=2mm, right=2mm, boxrule=1pt]
\footnotesize{
\textbf{Examples of LLM Responses to Design2SVA (FSM)}

\vspace{5pt}

{\normalsize \textbf{gpt-4-turbo (gpt-4-0125-preview):}}

\vspace{5pt}

\textbf{Attempt 1}
\begin{lstlisting}[language=Verilog]
logic [1:0] prev_state, curr_state;
always_ff @(posedge clk or negedge reset_) begin
    if (!reset_) begin
        prev_state <= S0;
    end else begin
        prev_state <= curr_state;
    end
end

assign curr_state = fsm_out;

assert property (@(posedge clk) disable iff (!reset_)
    (prev_state == S2 && curr_state == S0) |-> ##[1] (in_A != in_B)
);
\end{lstlisting}

\textbf{Attempt 2}
\begin{lstlisting}[language=Verilog]
logic [1:0] test_state, test_next_state;
assign test_state = fsm_out; // Assuming fsm_out reflects the current state
always_comb begin
    case(test_state)
        S0: test_next_state = S2;
        S1: test_next_state = S3;
        S2: test_next_state = ((in_A != in_B) < 'd0) ? S0 : S1;
        S3: test_next_state = S3; // Assuming S3 is a terminal state with no transition defined
        default: test_next_state = S0; // Default case to handle latches
    endcase
end

assert property (@(posedge clk) disable iff (!tb_reset)
    (test_state == S0) |-> (test_next_state == S2) &&
    (test_state == S1) |-> (test_next_state == S3) &&
    (test_state == S2) |-> ((in_A != in_B) < 'd0 ? (test_next_state == S0) : (test_next_state == S1))
);
\end{lstlisting}
}
\end{tcolorbox}
\noindent\begin{minipage}{\textwidth}
\captionof{figure}{Example responses from \texttt{gpt-4-turbo} for the FSM test instance shown in ~\ref{subsec:appendix_deisng2sva_examples}. 
Two attempts from the model are shown.}\label{box:design2sva_model_response_fsm_gpt-4-turbo}
\end{minipage}
\newpage

\begin{tcolorbox}[width=1.0\linewidth, halign=left, colframe=black, colback=white, boxsep=0.01mm, arc=1.5mm, left=2mm, right=2mm, boxrule=1pt]
\footnotesize{
\textbf{[Continued] Examples of LLM Responses to Design2SVA (FSM)}

\vspace{5pt}

{\normalsize \textbf{gpt-4o-05-13:}}

\vspace{5pt}

\textbf{Attempt 1}
\begin{lstlisting}[language=Verilog]
logic [1:0] fsm_state, fsm_next_state;
assign fsm_state = fsm_out;

assert property (@(posedge clk) disable iff (tb_reset)
    (fsm_state == S2) |-> ((in_A != in_B) ? (fsm_next_state == S0) : (fsm_next_state == S1))
);
\end{lstlisting}

\textbf{Attempt 2}
\begin{lstlisting}[language=Verilog]
logic [1:0] fsm_state, next_fsm_state;
assign fsm_state = fsm_out;

always_comb begin
    case(fsm_state)
        S0: next_fsm_state = S2;
        S1: next_fsm_state = S3;
        S2: begin
            if (((in_A != in_B) < 'd0)) begin
                next_fsm_state = S0;
            end
            else begin
                next_fsm_state = S1;
            end
        end
        S3: next_fsm_state = S3; // S3 is a terminal state
        default: next_fsm_state = S0; // default case to avoid latches
    endcase
end

assert property (@(posedge clk) disable iff (tb_reset)
    (fsm_state == S2) |-> (next_fsm_state == S0 || next_fsm_state == S1)
);
\end{lstlisting}
}
\end{tcolorbox}
\noindent\begin{minipage}{\textwidth}
\captionof{figure}{Example responses from \texttt{gpt-4o} for the FSM test instance shown in ~\ref{subsec:appendix_deisng2sva_examples}. 
Two attempts from the model are shown.}\label{box:design2sva_model_response_fsm_gpt4o}
\end{minipage}
\newpage

\begin{tcolorbox}[width=1.0\linewidth, halign=left, colframe=black, colback=white, boxsep=0.01mm, arc=1.5mm, left=2mm, right=2mm, boxrule=1pt]
\footnotesize{
\textbf{[Continued] Examples of LLM Responses to Design2SVA (FSM)}

\vspace{5pt}

{\normalsize \textbf{Mixtral-8x22B-instruct-v1:}}

\vspace{5pt}

\textbf{Attempt 1}
\begin{lstlisting}[language=Verilog]
logic [WIDTH-1:0] in_A_reg, in_B_reg;
always_ff @(posedge clk or negedge reset_) begin
    if (!reset_) begin
        in_A_reg <= '0;
        in_B_reg <= '0;
    end else begin
        in_A_reg <= in_A;
        in_B_reg <= in_B;
    end
end

assert property (@(posedge clk) disable iff (tb_reset)
    $rose(fsm_out == S0) |-> ##1 (in_A_reg != in_B_reg)
);
\end{lstlisting}

\textbf{Attempt 2}
\begin{lstlisting}[language=Verilog]
logic [WIDTH-1:0] a, b;
assign a = in_A;
assign b = in_B;

assert property (@(posedge clk) disable iff (tb_reset)
    $rose(state == S2) |-> (a == b) until (state == S0)
);
\end{lstlisting}
}
\end{tcolorbox}
\noindent\begin{minipage}{\textwidth}
\captionof{figure}{Example responses from \texttt{Mixtral-8x22B} for the FSM test instance shown in ~\ref{subsec:appendix_deisng2sva_examples}. 
Two attempts from the model are shown.}\label{box:design2sva_model_response_fsm_mixtral_8x22b}
\end{minipage}

\newpage

\end{document}